\documentclass[preprint,10pt]{elsarticle}
\usepackage{bm}
\usepackage{textcomp}
\usepackage{graphicx}
\usepackage{natbib}
\usepackage{geometry}
\usepackage{pifont}
\usepackage[nodots]{numcompress}
\journal{Journal of Quantitative Spectroscopy and Radiative Transfer}

\begin{document}
\begin{frontmatter}
\title{ Polarized line formation with $J$-state interference
in the presence of magnetic fields: A heuristic treatment
of collisional frequency redistribution}
\author[label1]{H. N. Smitha}
\ead{smithahn@iiap.res.in}
\author[label1]{K. N. Nagendra}
\ead{knn@iiap.res.in}
\author[label1]{M. Sampoorna}
\ead{sampoorna@iiap.res.in}
\author[label2,label3]{J. O. Stenflo}
\ead{stenflo@astro.phys.ethz.ch} 
\address[label1]{Indian Institute of Astrophysics, Koramangala, Bangalore 560034, India}
\address[label2]{Institute of Astronomy, ETH Zurich, CH-8093 \  Zurich, Switzerland}
\address[label3]{Istituto Ricerche Solari Locarno, Via Patocchi, 6605 Locarno-Monti, Switzerland}

\begin{abstract}
An expression for the partial frequency redistribution (PRD) matrix
for line scattering in a two-term atom, which includes the
$J$-state interference between its fine structure line components is derived.
The influence of  collisions (both elastic and inelastic) and
an external magnetic field on the
scattering process is taken into account. 
The lower term is assumed to be unpolarized and infinitely sharp.
The linear Zeeman regime in which the
Zeeman splitting is much smaller than the
fine structure splitting is considered.
The inelastic collision rates between 
the different levels are included in our treatment. 
{We account for the depolarization caused by
the collisions coupling the fine structure states of the upper 
term, but neglect the polarization transfer between the 
fine structure states}.
When the fine structure splitting goes to zero, 
we recover the redistribution matrix 
that represents the scattering on a two-level atom 
(which exhibits only $m$-state 
interference --- namely the Hanle effect). The way
in which the multipolar index of the scattering atom enters into 
the expression for the redistribution matrix through the collisional
branching ratios is discussed. 
The properties of the redistribution matrix are explored 
for a single scattering process for an $L=0 \to 1 \to 0$ scattering 
transition with $S=1/2$ (a hypothetical doublet
centered at 5000\,\AA\ and 5001\,\AA). 
Further, a method for solving the Hanle radiative 
transfer equation for a two-term atom in the presence of collisions,
PRD, and $J$-state interference is developed. The Stokes 
profiles emerging from an isothermal constant property medium
are computed.
\end{abstract}


\end{frontmatter}

\section{Introduction}
\label{intro}
The Solar spectrum is linearly polarized due to coherent scattering 
processes in the Sun's atmosphere. This linearly polarized spectrum is
as rich in spectral structures as the ordinary intensity spectrum, 
but it differs in appearance and information 
contents (see \citet{sk96,sk97}, \citet{josetal97}). It is therefore referred to as the 
``Second Solar Spectrum''. A weak magnetic field modifies this spectrum 
through the process of the Hanle effect. This makes the Second Solar Spectrum 
sensitive to a field strength regime that is inaccessible to the 
ordinary Zeeman effect. It therefore has the potential to 
significantly advance our understanding 
of the Sun's magnetism. In order to interpret the 
wealth of information imprinted 
in the Second Solar Spectrum, it is necessary to develop adequate theoretical 
tools that can later be used for the polarized line formation calculations.   

\citet{s94,s98} (hereafter S94 and S98 respectively) developed 
a classical theory for frequency-coherent scattering 
of polarized radiation in the presence of magnetic fields of 
arbitrary strength and orientation. This theory was generalized by 
\citet{bs99} (hereafter BS99) to include the effects of partial frequency 
redistribution (PRD) in the scattering process.
Their formulation was restricted to the 
rest frame of the atom. The transformation to the laboratory frame 
was presented in \citet{ms07a} (hereafter HZ1) for the special 
case of a $J=0 \to 1 \to 0$ scattering transition. Recently \citet{ms11} 
has generalized the classical PRD theory of BS99 to treat other types of 
atomic transitions with arbitrary $J$-quantum numbers. 

\begin{figure}
\centering
\includegraphics[height=3.2cm, width=9.0cm]{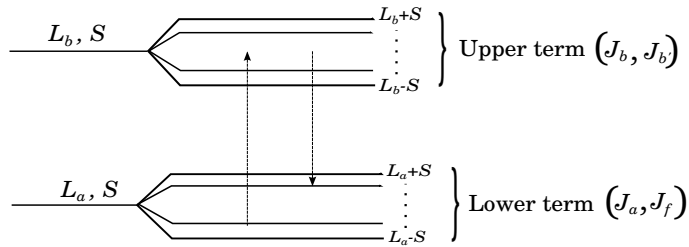}
\caption{{ Schematic} Level diagram of a two-term atom. The lower term is assumed to be
infinitely sharp whereas the upper term is both radiatively and collisionally
broadened.}
\label{level-diag}
\end{figure}

In \citet{p1} (hereafter P1) we 
derived the polarized PRD matrices for a two-term atom 
with an arbitrary $L_{a}\to L_{b}\to L_{a}$ 
scattering transition, taking into account the effects of $J$-state 
interference between the fine structure components of the 
split upper term $L_b$  (see Figure~{\ref{level-diag}}). 
However, these expressions  were limited to the collisionless 
regime. In the present 
paper we generalize the semi-classical theory of \citet{ms11} to include 
$J$-state interference for a two-term atom in the presence of collisions. 
In \citet{p3} (hereafter P3), a simpler version of this theory (see 
Section~{\ref{lab-redis}}) has already been applied 
to model the non-magnetic linear polarization observations of $J$-state
interference phenomena in the Cr\,{\sc i} triplet.

Collisions play a vital role in determining the polarization
properties of the scattered radiation. For the case of a two-level atom
with unpolarized lower level, \citet{osc72} developed the quantum theory 
of polarized scattering in a non-magnetic medium, including PRD effects. 
They describe in detail the role played by elastic and 
inelastic collisions. The effects of magnetic fields were considered in 
\citet{osc73}. An explicit form of the polarized PRD matrix for resonance 
scattering on a two-level atom was derived by \citet{dh88}, based on the 
work of \citet{osc72}, assuming that the lower level is unpolarized. 
Under the same assumption, a more elegant form of the PRD matrix for 
both the non-magnetic and magnetic cases was derived in pioneering 
papers by \citet[]{vb97a,vb97b,vb03} using the master equation theory. 
The equivalence between the QED theory of \citet{vb97b} and the 
semi-classical theory was demonstrated in \citet{ms07b} (hereafter HZ2)
 for a $J=0\to 1\to 0$ scattering transition, and in \citet{ms11} 
for an arbitrary $J_a\to J_b\to J_a$ scattering transition. An alternative 
PRD theory based on the concept of metalevels has been developed by 
\citet{landi97} for the collisionless case. This formulation can also
deal with $J$-state interference in the presence of magnetic fields.

In the present paper, starting from the Kramers-Heisenberg formula, we 
derive the expressions for the collisional PRD matrices including the 
effects of $J$-state interference for a two-term atom. 
The following assumptions are made:
\begin{enumerate}
\item Infinitely sharp lower term.
\item Unpolarized lower term.
\item Weak radiation field limit (i.e., stimulated emission is neglected 
in comparison with the spontaneous emission).
\item Hyperfine structure is neglected.
\item The effects of inelastic collisions that couple the 
fine structure states are treated approximately (see below).
\item The depolarizing elastic collisions that couple $m$-states 
belonging to a given fine structure state $J_b$ are taken into account,
but are assumed to be independent of the $J$-quantum number {for the sake
of mathematical simplicity}.
\item We restrict our attention to the linear Zeeman regime of magnetic
field strengths.
\end{enumerate}

The assumption of an unpolarized lower term is made for the sake of 
mathematical simplicity, but can often be justified when the
lower term represents the ground state of the atom. In the stellar
atmospheric conditions the ground state is generally two orders of magnitude
more long lived than the 
excited state, which makes it correspondingly much harder
for any ground state 
polarization to survive collisional and magnetic depolarization, 
as compared with the excited states
(see \citet{kvb02}).
We also ignore the induced emission, because in scattering problems 
it acts as a negative absorption and only affects the radiation in the
exact forward direction (scattering angle exactly zero). The induced 
emission probability is nearly three orders of magnitude smaller
than the spontaneous emission probability (see \citet{kvb02}).

The inelastic collisions between
the upper and lower terms are treated exactly while the inelastic
collisions between the upper fine structure states are treated approximately.
The inelastic collisions between the upper fine structure
states (denoted by $\Gamma_{IJ_bJ_{b^\prime}}$) manifest themselves
in two different ways, (i) through a depolarization of state $J_b$ and 
(ii) through a transfer of { alignment and orientation} between $J_b$ and
$J_{b^\prime}$. 

Since the colliding particles are isotropically distributed 
around the radiating atom, they destroy the alignment and thereby
depolarize the levels. Therefore the inelastic collisions  that take the atom
away from the state $J_b$ always depolarize $J_b$. They also contribute to 
the inverse lifetime of $J_b$ under consideration. In this paper we take
into account such inelastic collisions between the fine structure states
$J_b$ and $J_{b^\prime}$ by adding these inelastic collision rates 
$(\Gamma_{IJ_bJ_{b^\prime}})$ to the inelastic collision rate 
$\Gamma_{IJ_bJ_f}$ (where $f$ is the final state).
The depolarizing effects of these inelastic collisions are similar to the 
depolarizing effects of elastic collisions. Thus we merge these two 
effects and define a common damping rate $\gamma_b$ for the $J_b$ state.

The inelastic collisions between polarized fine structure states can lead 
to a transfer of {alignment and orientation} between them {(hereafter
referred to as the transfer of polarization)}. This is similar to optical pumping
by radiative transitions. The only difference is that the radiative transitions
between the fine structure states $J_b$ and $J_{b^\prime}$ 
are not allowed. Taking account of such transfer rates
caused by inelastic collisions actually involves formulating
the statistical equilibrium equations for the concerned states
including the atomic polarization of the various states. This is outside the
scope of our present paper.
A formulation of statistical equilibrium equation including these collisions
but neglecting the redistribution effects in scattering has
been presented in \citet{kerk02} and \citet{kvb02}.
They derive the expressions to calculate these rates taking examples of 
few atomic systems of relevance to the analysis of the
second solar spectrum.
Our present treatment of inelastic collisions is 
basically heuristic and only takes
into account the depolarizing effects of $\Gamma_{IJ_bJ_{b^\prime}}$. 

The frequency redistribution function
that describes the effect of collisions in unpolarized radiative
transfer is the well known type-III 
(or $R^{\rm III}$) function of \citet{hum62}. Here we describe the
matrix generalizations of this standard collisional redistribution
function,  brought about by the magnetic fields and the $J$-state
interference. Using the method described in Appendix C of HZ2, we 
rewrite the PRD matrices in terms 
of the irreducible spherical tensors for polarimetry. We discuss in 
detail the procedure to 
identify the multipolar index $K$, which needs to be assigned to the branching 
ratios that govern the effect of the depolarizing collisions. We
illustrate  the effects of collisions on the Stokes 
$(I,Q/I,U/I,V/I)$ profiles of the scattered radiation for the 
90\textdegree \ single scattering case. Then we present the 
technique of incorporating this Hanle redistribution
matrix for the two-term atom
into the polarized radiative transfer equation, and solve it for an isothermal
constant property atmospheric slab. In the collisionless case, the relevant 
redistribution matrix derived in P1 was incorporated into the transfer equation
 in \citet{p2} (hereafter P2) and solved for a constant property isothermal 
media in the absence of a magnetic field.
The same method of solution presented in P2 is also used here, but including
the collisional redistribution in the presence of a magnetic field. 

In Section~{\ref{hanle-zeeman}} we derive the elements of the ensemble averaged
coherency matrix both in the atomic and laboratory frames for a 
$L_a \to L_b \to L_a$
scattering transition taking into account the elastic collisions. In 
Section~{\ref{hz-tkq}} we express the type-III redistribution matrix 
in terms of the irreducible spherical tensors both for the non-magnetic
and magnetic cases. The important question of identifying the multipolar
index $K$ that describes the transfer of angular momentum in 
a scattering event affected by the depolarizing collisions is discussed in 
detail. The laboratory frame expression for the collisional redistribution
matrix is also derived in this section. The procedure to incorporate this
redistribution matrix into the polarized radiative transfer equation for both
the magnetic and non-magnetic cases is discussed in Section~{\ref{rt-gen}}.
The Stokes profiles resulting from 
a single $90^\circ$ scattering event and from multiple scattering in 
an isothermal atmospheric slab are presented in Section~{\ref{results}}. 
Concluding remarks are given in Section~\ref{conclu}.
{ Finally, in Appendix~A we give the expressions for the magnetic 
redistribution
functions of type-III.}

\section{An approximate treatment of collisions including the $J$-state
 interference}
\label{hanle-zeeman}
\subsection{A semi-classical formulation for the polarized PRD matrix}

The Mueller matrix ${\bf M}$ that describes the transformation from the incident 
to the scattered Stokes vector is given by
\begin{equation}
{\bf M}={\bf T}{\bf W}{\bf T}^{-1},
\label{mueller_mat}
\end{equation}
where ${\bf T}$ and ${\bf T}^{-1}$ are purely 
mathematical transformation matrices. Their explicit forms are given 
in Equation~(9) of S98. 
The ${\bf W}$-matrix is defined in Equation~(7)
of P1. The elements of this matrix contain the bilinear products 
of the complex probability amplitude $w_{\alpha\beta}(J_f\mu_fJ_a\mu_a)$.
These amplitudes for transition from an initial state $a$
to final state $f$ via all intermediate states $b$ are given by the
Kramers-Heisenberg formula (see S98 and \citet{ms11} for historical
accounts) as
\begin{equation}
 w_{\alpha\beta} \sim \sum_{b}\frac{\left\langle 
f|{\bf{r\  .\  e_{\alpha}}}|b\right\rangle \left\langle 
b|{\bf{r\  .\  e_{\beta}}}|a\right\rangle}{\omega_{bf}-\omega-{\rm i}\gamma_b/2},
\label{k-h}
\end{equation}
where $\omega = 2\pi \xi$ is the angular frequency of the 
scattered radiation in the atomic rest frame, $\hbar\omega_{bf}$
is the energy difference between the excited and 
final states, and $\gamma_b$ is the damping constant that accounts for
the broadening of the excited state $b$, while the 
initial and the final states are assumed to be infinitely sharp. 
The damping parameter is assumed to be the same for all the magnetic
substates of the excited state. The matrix elements appearing in 
Equation~(\ref{k-h}) can be expanded using the Wigner-Eckart theorem
as
\begin{eqnarray}
&&w_{\alpha\beta}(J_f\mu_fJ_a\mu_a) \sim \sum_{J_b\mu_b} (-1)^{q-q^\prime} 
 \sqrt{(2J_a+1)(2J_f+1)} (2J_b+1)(2L_a+1) 
\left\lbrace 
\begin{array}{ccc}
L_a & L_b & 1\\
J_b & J_f & S \\
\end{array}
\right\rbrace \nonumber \\ && \times 
\left\lbrace 
\begin{array}{ccc}
L_a & L_b & 1\\
J_b & J_a & S \\
\end{array}
\right\rbrace
\left(
\begin{array}{ccc}
J_b & J_a & 1 \\
-\mu_b & \mu_a& -q^\prime \\
\end{array}
\right)
\left(
\begin{array}{ccc}
J_b & J_f & 1 \\
-\mu_b & \mu_f & -q \\
\end{array}
\right)
\Phi_{\gamma_b}(\nu_{J_b\mu_bJ_f\mu_f} - \xi)
 \varepsilon^{\alpha\ast}_q \varepsilon^{\beta}_{q^\prime},
\label{khform_w}
\end{eqnarray}
where $\mu_b$ represents the magnetic substates of the 
upper state $b$ with total angular momentum quantum number $J_b$, orbital
 angular momentum quantum number $L_b$, and spin $S$. The quantities 
$J_a$ and $J_f$ 
are respectively the total angular momentum quantum numbers of the  
initial and final states $a$ and $f$ with orbital angular momentum 
quantum number $L_a$, and magnetic substates $\mu_a$ and 
$\mu_f$. The quantities $\varepsilon$ are the geometrical factors (see 
Equations~(2) and (27) of S98), with $\alpha$ and $\beta$ 
denoting the outgoing and incoming radiation, respectively. In 
Equation~(\ref{khform_w}), $q=\mu_f-\mu_b$ and $q^\prime=\mu_a - \mu_b$. 
{ In the rest of the paper we denote the indices as follows for the sake of
convenience 
\begin{eqnarray}
&&J_b=b,\ J_a=a,\ J_f=f; \quad J_b\mu_b=b_m,\ J_a\mu_a=
a_m,\ J_f\mu_f=f_m,\nonumber \\ &&
\ \ \ \ \ \ \ \ \ \ \ \ \ \ J_{b^\prime}\mu_{b^\prime}=b'_m,\  
J_{b}\mu_{b^{\prime\prime}}=b''_m,\  J_{b^\prime}\mu_{b^{\prime\prime\prime}}
=b'''_m.
\end{eqnarray} }
The frequency-normalized profile function is given by
\begin{equation}
\Phi_{\gamma_b}(\nu_{b_mf_m}-\xi) =\frac{1/(\pi\rm{i})}
{\nu_{b_mf_m}-\xi-{\rm i}\gamma_{b}/(4\pi)}\ \  {\rm{with}}\ \ \ \nu_{b_mf_m} = 
\nu_{bf} + (g_b \mu_b - g_f \mu_f) \nu_L.
\label{prof_func}
\end{equation}
Here $h\nu_{bf}$ is the energy difference between the upper 
($J_b$) and lower ($J_f$) states in the absence of magnetic fields, 
$g_{b}$, $g_{f}$ are the Land\'e factors of these states, and $\nu_L$ is 
the Larmor
 frequency. Equation~(\ref{khform_w}) refers to the case of 
frequency-coherent scattering in the atomic rest frame. 

The phenomenological extension of Equation~(\ref{khform_w}) to the case of PRD
 is achieved by treating 
each radiative emission transition between magnetic substates $\mu_b$ 
and $\mu_f$ in terms of a damped oscillator that is truncated by 
collisions (see HZ1). In other words, in 
Equation~(\ref{khform_w})  we make the 
following replacement for the profile function\,:
\begin{equation}
\Phi_{\gamma_b}(\nu_{b_mf_m}-\xi) \longrightarrow 
(\tilde r_{b_m})_{a_mf_m},
\label{phen_extn_toprd}
\end{equation}
where the Fourier-transformed solution of the time-dependent 
oscillator equation is given by (see BS99)
\begin{eqnarray}
(\tilde r_{b_m})_{a_mf_m}= (\tilde r^{\rm stat}_{b_m})_{a_mf_m}
+\ C\ (\tilde r^{\rm trans}_{b_m})_{a_mf_m}.
\label{rtilde}
\end{eqnarray}
Here we have omitted the unimportant phase factor, as it vanishes 
in the bilinear product \\ $(\tilde r_{b_m})_{a_mf_m}
 (\tilde r^\ast_{{b^\prime}_m})_{a_mf_m}$. 
The constant $C$ in Equation~(\ref{rtilde}) defines the relative
amplitudes of the stationary and the transitory parts of the solution,
which are given by 
\begin{eqnarray}
(\tilde r^{\rm stat}_{b_m})_{a_mf_m} = 
\Phi_{\gamma_b}(\nu_{b_ma_m}-\xi^\prime)
\delta(\xi-\xi^\prime-\nu_{a_mf_m}),
\label{rtilde_stat}
\end{eqnarray}
\begin{eqnarray}
(\tilde r^{\rm trans}_{b_m})_{a_mf_m} =
\Phi_{\gamma_b}(\nu_{b_ma_m}-\xi^\prime) 
\Phi_{\gamma_b}(\nu_{b_mf_m}-\xi)
\left[1-{\rm e}^{-{\rm i}(\omega_{b_mf_m}-{\rm i}
\gamma_{b}/2-\omega)t_c}\right]. 
\label{rtilde_trans}
\end{eqnarray}
Here $\xi^\prime$ denotes the frequency of the 
incoming photon in the atomic rest frame, $t_c$ is the time between
two successive  collisions,
$\omega_{b_mf_m}=2\pi\nu_{b_mf_m}$. The profile function 
$\Phi_{\gamma_b}(\nu_{b_ma_m}-\xi^\prime)$ is given
by Equation~(\ref{prof_func}) with $\xi$ replaced by $\xi^{\prime}$,
while $\nu_{b_mf_m}$ is replaced by $\nu_{b_ma_m}$
that is defined similar to Equation~(\ref{prof_func}).
In Equation~(\ref{rtilde_stat}), $\nu_{a_mf_m}$ appearing in
the delta function is demanded by energy conservation 
(see Equation~(9.10) of S94), and is given by
\begin{equation}
\nu_{a_mf_m} = \nu_{af} + (g_a\mu_a-g_f\mu_f)\nu_L,
\label{nu-af}
\end{equation}
where $h\nu_{af}$ is the energy difference between the states $J_a$ and $J_f$
 in the absence of a magnetic field.
\subsection{Coherency matrix in the atomic rest frame}
\label{atom-frame}
The  elements of the ensemble averaged coherency matrix 
$\langle \tilde r_{b_m} \tilde r^\ast_{{b^\prime}_m} \rangle_{a_mf_m}$ 
can be derived starting from Equations~(\ref{rtilde_stat}) and 
(\ref{rtilde_trans}), 
 applying the same steps that are described in detail in BS99. 
These elements are contained
 in the bilinear product $w_{\alpha\beta}(f_ma_m)
w^\ast_{\alpha^\prime\beta^\prime}(f_ma_m)$. In the atomic rest frame ensemble averaged 
coherency matrix elements are given by
\begin{eqnarray}
&&\langle \tilde r_{b_m} \tilde r^\ast_{{b^\prime}_m} 
\rangle_{a_mf_m} = A_{b{b^\prime}} \cos\beta_{{b^\prime}_mb_m}\  
{\rm e}^{{\rm i}\beta_{{b^\prime}_mb_m}}\ 
\Phi^ { {{\gamma}_{bb^\prime}+\gamma_c}}_{b_m{b^\prime}_ma_m}(\xi^\prime)
\delta(\xi-\xi^\prime-\nu_{a_mf_m}) \nonumber\\ && +
B_{b{b^\prime}} \cos\beta_{{b^\prime}_m
b_m}\cos\alpha_{{b^\prime}_mb_m}
{\rm e}^{{\rm i}(\beta_{{b^\prime}_mb_m}+
\alpha_{{b^\prime}_mb_m})} 
\Phi^{{{\gamma}}_{bb^\prime}+\gamma_c}_{b_m{b^\prime}_ma_m}(\xi^\prime)
\Phi^{{{\gamma}}_{bb^\prime}+\gamma_c}_{b_m{b^\prime}_mf_m}(\xi),
\label{coherency_mat_atom}
\end{eqnarray}
where the angles $\beta_{{b^\prime}_mb_m}$ and 
$\alpha_{{b^\prime}_mb_m}$ (arising due to the combined effects of the
 $J$-state and $m$-state interferences) are defined respectively by 
\begin{eqnarray}
\tan \beta_{{b^\prime}_mb_m} = 
{\omega_{{b^\prime}b}+(g_{b^\prime}\mu_{b^\prime}-g_b\mu_b) \omega_L \over 
{{\gamma}_{{b^\prime}b}+\gamma_c}};\quad
\tan \alpha_{{b^\prime}_mb_m} = 
{\omega_{{b^\prime}b}+(g_{b^\prime}\mu_{b^\prime}-g_b\mu_b) \omega_L \over 
{{\gamma}_{{b^\prime}b}+\gamma_c/2}},
\label{angle_alpha-beta}
\end{eqnarray}
with  ${\gamma}_{bb^\prime}$ 
given by
\begin{equation}
 {\gamma}_{b{b^\prime}}=
\frac{\gamma_{b}+\gamma_{{b^\prime}}}{2}={\gamma}_{{b^\prime}b}.
\end{equation}
Here $\gamma_c$ is the collisional damping constant, while  
$\hbar \omega_{{b^\prime}b}$ is the energy difference between
the $J_{b^\prime}$ and $J_b$ states in the absence of a magnetic
field.  The elastic collisional rates are in general different for 
each fine structure component ($J_b$) of the upper term. 
{However, for simplicity we assume them to be 
independent of the $J$-quantum numbers.}

$A_{b{b^\prime}}$ and $B_{b{b^\prime}}$ are the branching ratios for 
a two-term atom. 
The explicit expressions for them will be defined later in 
Section~{\ref{multi-k}}.

Like in P1, we limit the treatment to the
linear Zeeman regime, in which the Zeeman splitting is much
smaller than the fine structure splitting. When
$J_{b}\ne J_{b^\prime}$ the contributions from the second terms
with $\omega_{L}$ in Equation~(\ref{angle_alpha-beta}) to the
angles $\beta_{{b^\prime}_mb_m}$ and
$\alpha_{{b^\prime}_mb_m}$ can therefore be ignored, because they
are insignificant in comparison with the first terms.
The classical generalized profile function is defined as

\begin{equation}
\Phi^{{\gamma}_{bb^\prime}}_{b_m{b^\prime}_mf_m}(\xi)={1\over 2} 
\left[ \Phi_{\gamma_b}(\nu_{b_mf_m}-\xi) + 
\Phi^\ast_{\gamma_{b^\prime}}(\nu_{{b^\prime}_mf_m}-\xi)\right],
\label{gen_prof}
\end{equation}
in the same way as in BS99.
\subsection{Coherency matrix in the laboratory frame for 
type-III redistribution}
\label{lab-frame}
We transform Equation~(\ref{coherency_mat_atom}) to the 
laboratory frame using the same steps as described 
in Section~2.2 of HZ2. Thus the ensemble averaged coherency matrix in the 
laboratory frame is given by
\begin{eqnarray}
&&\!\!\!\!\!\!\!\!
\langle \tilde r_{b_m} \tilde r^\ast_{{b^\prime}_m} 
\rangle_{a_mf_m} = A_{b{b^\prime}} 
\cos\beta_{{b^\prime}_mb_m} 
{\rm e}^{{\rm i}\beta_{{b^\prime}_mb_m}}
\Big[(h^{\rm II}_{b_m,{b^\prime}_m})_{a_mf_m}
+{\rm i}(f^{\rm II}_{b_m,{b^\prime}_m})_{a_mf_m}
\Big]\ \nonumber \\ &&
+ B_{b{b^\prime}} \cos\beta_{{b^\prime}_mb_m}
\cos\alpha_{{b^\prime}_mb_m}
{\rm e}^{{\rm i}(\beta_{{b^\prime}_mb_m}+
\alpha_{{b^\prime}_mb_m})} 
\Big[h^{\rm III}_{b_ma_m,{b^\prime}_mf_m}
+ {\rm i} f^{\rm III}_{b_ma_m,{b^\prime}_mf_m}
\Big] .
\label{coherency_mat_lab}
\end{eqnarray}
The various auxiliary quantities for type-II redistribution are defined in
 Section~3 of P1. 
Hence we do not repeat them here. Hereafter we confine our attention to the 
collisional 
redistribution (type-III). The corresponding derivation for pure
radiative (collisionless) redistribution (type-II) are given in P1.
The auxiliary quantities 
for type-III redistribution that appear in
Equation~(\ref{coherency_mat_lab}) are defined by 
\begin{eqnarray}
h^{\rm III}_{b_ma_m,{b^\prime}_mf_m} &=& 
{\frac{1}{4}}\,\Bigg[
R^{\rm III,\,HH}_{{b^\prime}_ma_m,{b^\prime}_mf_m}+
R^{\rm III,\,HH}_{{b^\prime}_ma_m,b_mf_m}+
R^{\rm III,\,HH}_{b_ma_m,{b^\prime}_mf_m}+
R^{\rm III,\,HH}_{b_ma_m,b_mf_m}\Bigg] \nonumber  \\ && +
{\frac{\rm i}{4}}\,\Bigg[
R^{\rm III,\,FH}_{{b^\prime}_ma_m,{b^\prime}_mf_m}+
R^{\rm III,\,FH}_{{b^\prime}_ma_m,b_mf_m} -
R^{\rm III,\,FH}_{b_ma_m,{b^\prime}_mf_m} -
R^{\rm III,\,FH}_{b_ma_m,b_mf_m}\Bigg].
\label{h3mubmubprimemuf}
\end{eqnarray}
Similarly we have
\begin{eqnarray}
f^{\rm III}_{b_ma_m,{b^\prime}_mf_m} &=& 
{\frac{1}{4}}\,\Bigg[R^{\rm III,\,HF}_{{b^\prime}_ma_m,{b^\prime}_mf_m}-
R^{\rm III,\,HF}_{{b^\prime}_ma_m,b_mf_m}+R^{\rm III,\,HF}_
{b_ma_m,{b^\prime}_mf_m}-R^{\rm III,\,HF}_{b_ma_m,b_mf_m}\Bigg]
\nonumber \\ && +
{\frac{\rm i}{4}}\,\Bigg[R^{\rm III,\,FF}_{{b^\prime}_ma_m,{b^\prime}_mf_m}-
R^{\rm III,\,FF}_{{b^\prime}_ma_m,b_mf_m} -
R^{\rm III,\,FF}_{b_ma_m,{b^\prime}_mf_m}+
R^{\rm III,\,FF}_{b_ma_m,b_mf_m}\Bigg].
\label{f3mubmubprimemuf}
\end{eqnarray}
{The magnetic redistribution functions of type-III appearing 
in the above equations 
are defined in {\ref{appendix}}}.
\section{The PRD matrix expressed in terms of irreducible tensors}
\label{hz-tkq}
The importance of expressing the PRD matrices in terms of the irreducible
 spherical tensors 
introduced by \citet{landi84} has been discussed in P1. The definition
 and properties of 
irreducible spherical tensors are described in detail in 
\citet{ll04} (hereafter LL04).
The way to incorporate these tensors in the 
analytic form of the PRD matrix derived from a semi-classical
approach has been described in HZ2 
(see also Section~4 of P1). Applying the same method we have obtained an
expression for the type-III redistribution matrix in terms of irreducible 
spherical tensors. The case of the 
type-II redistribution matrix has been discussed in Section~4 of P1.

As in P1 we now express the 
type-III PRD matrix derived in Section~{\ref{hanle-zeeman}} in terms of
 ${\mathcal T}^K_Q(i,{\bm n})$, where 
$i=0,1,2,3$, and $K=0,1,2$ with $-K\le Q\le +K$.
Following the same procedure as discussed in Section~4 of P1, the matrix 
$T^{\rm S}_{\mu\nu,\rho\sigma}$ of Equation~(22) in P1, which describes
the transformation of the elements of the coherency matrix, can be
written in the atomic rest frame as 
\begin{eqnarray}
&&\!\!\!\!\!\!\!\!\!T^{\rm S}_{\mu\nu,\rho\sigma}(\xi,{\bm n};\xi^\prime,
{\bm n}^\prime,{\bm B}) 
= (2L_a+1)^{2} \sum_{a_mf_mb_m{b^\prime}_m} G\ Z_6 Z_3\ \ 
(-1)^{q-q^\prime+q^{\prime\prime}-q^{\prime\prime\prime}}
{\mathcal E}^{\rm S}_{qq^{\prime\prime}}
(\mu,\nu,{\bm n}){\mathcal E}^
{\rm S}_{q^{\prime\prime\prime}
q^\prime} 
(\sigma,\rho,{\bm n}^\prime)
\nonumber \\ && \times
\Big\{A_{b{b^\prime}} \cos\beta_{{b^\prime}_mb_m}\  
{\rm e}^{{\rm i}\beta_{{b^\prime}_mb_m}}\ 
\Phi^ { {{\gamma}_{bb^\prime}+\gamma_c}}_{b_m{b^\prime}_ma_m}(\xi^\prime)
\delta(\xi-\xi^\prime-\nu_{a_mf_m}) \nonumber\\ && +
B_{b{b^\prime}} \cos\beta_{{b^\prime}_m
b_m}\cos\alpha_{{b^\prime}_mb_m}
{\rm e}^{{\rm i}(\beta_{{b^\prime}_mb_m}+
\alpha_{{b^\prime}_mb_m})} 
\Phi^{{{\gamma}}_{bb^\prime}+\gamma_c}_{b_m{b^\prime}_ma_m}(\xi^\prime)
\Phi^{{{\gamma}}_{bb^\prime}+\gamma_c}_{b_m{b^\prime}_mf_m}(\xi)\Big\},
\label{ts_mat_prd}
\end{eqnarray}
where
\begin{equation}
G=(2J_a+1)(2J_f+1)(2J_b+1)(2J_{b^\prime}+1),
\end{equation}
\begin{equation}
Z_{6}=\left\lbrace 
\begin{array}{ccc}
L_a & L_b & 1\\
J_b & J_f & S \\
\end{array}
\right\rbrace
\left\lbrace 
\begin{array}{ccc}
L_a & L_b & 1\\
J_b & J_a & S \\
\end{array}
\right\rbrace
\left\lbrace 
\begin{array}{ccc}
L_a & L_b & 1\\
J_{b^\prime} & J_f & S \\
\end{array}
\right\rbrace
\left\lbrace 
\begin{array}{ccc}
L_a & L_b & 1\\
J_{b^\prime} & J_a & S \\
\end{array}
\right\rbrace, 
\end{equation}
and
\begin{equation}
 Z_3= \left(
\begin{array}{ccc}
J_b & J_a & 1 \\
-\mu_b & \mu_a& -q^\prime \\
\end{array}
\right)
\left(
\begin{array}{ccc}
J_b & J_f & 1 \\
-\mu_b & \mu_f & -q \\
\end{array}
\right) 
\left(
\begin{array}{ccc}
J_{b^\prime} & J_a & 1 \\
-\mu_{b^\prime} & \mu_a& -q^{\prime\prime\prime} \\
\end{array}
\right)
\left(
\begin{array}{ccc}
J_{b^\prime} & J_f & 1 \\
-\mu_{b^\prime} & \mu_f & -q^{\prime\prime} \\
\end{array}
\right).
\end{equation}

In Equation~(\ref{ts_mat_prd}), ${\mathcal E}^{\rm S}_{qq^{\prime\prime}}
(\mu,\nu,{\bm n})$ is a 
reducible spherical tensor. After transforming to the 
Stokes formalism (see Section~4 of P1), the redistribution matrix for $J$-state
interference can be written in symbolic form as
\begin{equation}
{\bf R}_{ij}(\xi,{\bm n}; \xi^\prime,{\bm n}^\prime,{\bm B}) = 
{\bf R}^{\rm II}_{ij}(\xi,{\bm n}; \xi^\prime,{\bm n}^\prime,{\bm B}) + 
{\bf R}^{\rm III}_{ij}(\xi,{\bm n}; \xi^\prime,{\bm n}^\prime,{\bm B}),  
\label{phen_rm}
\end{equation}
where the pure radiative part of the redistribution matrix is given by  branching 
ratio $A_{b{b^\prime}}$ times Equation~(25) of P1, and the collisional frequency 
redistribution is taken into account through
\begin{eqnarray}
&&{\bf R}^{\rm III}_{ij}(\xi,{\bm n}; \xi^\prime,{\bm n}^\prime,{\bm B})=
{ 2\over 3}(2L_a+1)^{2}
\sum_{K^\prime K^{\prime\prime}Qafb{b^\prime}}\  G\  Z_6\  B_{b{b^\prime}}
\sqrt{(2K^\prime+1)(2K^{\prime\prime}+1)} 
\nonumber \\ &&\times
\bigg\{\sum_{\mu_a\mu_f\mu_b\mu_{b^\prime}} Z_3\ 
(-1)^{q^{\prime\prime}+q^\prime+Q} 
\left(
\begin{array}{ccc}
1 & 1 & K^{\prime\prime} \\
q & -q^{\prime\prime} & Q \\
\end{array}
\right)
\left(
\begin{array}{ccc}
1 & 1 & K^{\prime} \\
q^{\prime\prime\prime} & -q^{\prime} & -Q \\
\end{array}
\right)
\nonumber \\ &&\times
{1\over 4} \left[\Phi_{\gamma_{b}+\gamma_c}(\nu_{b_ma_m}-\xi^\prime)
+\Phi^{\ast}_{\gamma_{b^\prime}+\gamma_c}(\nu_{{b^\prime}_m
a_m}-\xi^\prime)\right]
\left[\Phi_{\gamma_{b}+\gamma_c}(\nu_{b_mf_m}-\xi)              
+\Phi^\ast_{\gamma_{b^\prime}+\gamma_c}
(\nu_{{b^\prime}_mf_m}-\xi)\right]\nonumber \\ && \times
{\cos}\beta_{{b^\prime}_mb_m}\  
{\cos}\alpha_{{b^\prime}_mb_m}\ {\rm e}^{{\rm i}(\beta_{{b^\prime}_mb_m}+
\alpha_{{b^\prime}_mb_m})}
\bigg\} 
(-1)^Q{\mathcal T}^{K^{\prime\prime}}_Q(i,{\bm n})
{\mathcal T}^{K^{\prime}}_{-Q}(j,{\bm n}^\prime). 
\label{pehn_r3-1}
\end{eqnarray}
Note that in the formal expression for ${\bf R}$ 
the branching ratios are built into the ${\bf R}^{\rm II}$ 
and ${\bf R}^{\rm III}$ components. As the collisional branching 
ratio $B_{b{b^\prime}}$ depends on index $K$, our next task is to
determine the explicit form of this dependence. This will be done in the next
subsection.

\subsection{ Identification and physical significance of the multipolar
 index $K$ in the 
collisional branching ratios}
\label{multi-k}
It is well known that the spherical 
unit vectors form a natural basis to decouple the classical oscillator 
equation. \citet{fn57} suggested that a convenient 
basis to be used when dealing with scattering problems in quantum mechanics, 
is the irreducible tensorial { basis }
instead of the standard $|JM\rangle$ basis of Hilbert space. 
This is due to the fact that {irreducible tensors} transform under 
co-ordinate rotations like the spherical harmonics ($Y_{lm}$) and are thus 
suited for a study of rotationally invariant processes. With irreducible 
tensorial operators one can express the scattering matrix such that it 
formally looks the same in the magnetic (with the polar $z$-axis along ${\bm B}$) 
and the atmospheric (with the polar $z$-axis along the atmospheric normal) 
reference frames. This is the advantage of going to the irreducible tensorial 
basis 
(hereafter called the $KQ$ basis). A more detailed historical background 
for the irreducible tensorial operators is given in 
\citet{sbs97}. 

Thus the geometrical factors associated with the scattering problem, and 
also the density matrix for the atomic levels in question, should be transformed
 to 
the $KQ$ basis. The transformation of the geometrical factors to the $KQ$ basis 
is described in Chapter 5 of LL04 and is used in HZ2. The 
density matrix is first written in the standard $|JM\rangle$ basis and then 
transformed to the $KQ$ basis (see Equation~(3.97) of LL04), which is then 
called `multipole moments' of the density matrix, or `irreducible statistical 
tensors'. In the case of the radiation field, the multipole index $K$ has the 
following interpretation\,: $K=0$ means isotropic scattering, $K=1$ is 
related to the circular polarization, while $K=2$ is related to the linear 
polarization. In the case of the atomic levels, $K=0$ represents the population 
of the level under consideration, $K=1$ is related to the orientation of the 
atom, while $K=2$ is related to the alignment of the atom (this physical 
interpretation can be found in pp. 128 and 129 of LL04, Section 10.4 of S94, 
and in\citet{jtb01})

In the case of radiation field an irreducible tensor 
${\mathcal T}^K_Q(i,{\bm n})$ is constructed by forming a suitable linear combination 
of the direct product of two geometrical factors. Since geometrical factors 
basically contain unit polarization vectors of rank one, their direct product 
represents a second rank tensor, with $K$ taking values 0, 1, and 2. Note that
 these 
values of $K$ can also be obtained through angular momentum addition 
of two tensors of rank 1. In the case of the density matrix of the atom, 
the value of $K$ is determined by the addition of angular momenta $J$ and 
$J^\prime$. For example for a two-level atom with unpolarized ground level 
the value of $K$ relating to the statistical tensor of the upper level 
is given by angular momentum addition of $J_b$ and $J_{b^{\prime}}$. Further,
 $Q$ takes 
values $-K$ to $+K$ in steps of one, and is related to 
the magnetic quantum numbers of the upper level. 

We can denote $K^\prime$ as the multipole component of the incident radiation 
field, $K$ as the multipole moment of the upper level of the atom, and 
$K^{\prime\prime}$ as the multipole component of the scattered radiation. 
The scattering process can be understood as a transfer of the $K^\prime$ multipole 
component of the incident radiation to the $K$ multipole moment of the 
atom's upper level through an absorption process, followed by a transfer of the 
$K$ multipole moment of the atom's upper level to the $K^{\prime\prime}$ 
multipole component of the scattered radiation through spontaneous emission. 
The depolarizing collisions that  
govern the branching ratios and the magnetic field that governs the
Hanle angles affect the upper  
level of the atom directly and modify the $K$ multipole moment of the
atom, but they influence the scattered radiation 
only indirectly, through spontaneous emission from the level
that has been directly 
affected. Thus it is the $K$ index of the upper 
level of the atom that needs to be assigned to the branching ratios and the 
Hanle angles, and not the multipole component of the incident or the scattered 
radiation. In the absence of magnetic fields or in the presence of weak 
magnetic fields (Hanle effect), $K^\prime=K=K^{\prime\prime}$. In the presence 
of a magnetic field of arbitrary strength (Hanle-Zeeman regime) all 
the three $K$'s are 
distinct. This is due to the distinction preserved through the 
profile functions, which become different for the different Zeeman 
components. However, in weakly 
magnetic cases (when the Zeeman splitting is much smaller than the
effective line width) the distinction is so small that it can be ignored.

From the above discussion it is clear that for a correct identification of $K$ 
for the branching ratio we need to know the density matrix of the upper 
level in the $KQ$ basis. Since the density matrix does not appear directly in 
the Kramers-Heisenberg approach that we use, we need to indirectly identify 
$K$ either by drawing analogy with the density matrix theory, or by using a 
suitably defined quantum generalized profile function. 
Such a function was defined by \citet{lan91} for the special case of a two-level 
atom (without $J$-state interference). The multipole moment $K$ of
 the upper level is built into this function through 
the third $3-j$ symbol appearing in the following definition\,:
\begin{eqnarray}
&&\Phi^{K,K^\prime}_Q(J_a,J_b;\xi^\prime) =
\sqrt{3(2J_b+1)(2K+1)(2K^\prime+1)}
\sum_{\mu_b\mu_{b^\prime} \mu_a pp^\prime} (-1)^{J_b-\mu_a-1+Q}\nonumber \\ &&\times
\left(
\begin{array}{ccc}
J_b & J_a & 1 \\
-\mu_b & \mu_a & p \\
\end{array}
\right)
\left(
\begin{array}{ccc}
J_b & J_a & 1 \\
-\mu_{b^\prime} & \mu_a & p^{\prime} \\
\end{array}
\right)
\left(
\begin{array}{ccc}
J_b & K & J_b \\
-\mu_b & Q & \mu_{b^\prime} \\
\end{array}
\right)
\left(
\begin{array}{ccc}
1 & 1 & K^\prime \\
-p & p^{\prime} & Q \\
\end{array}
\right)\nonumber \\ &&\times
{1\over 2} \left[\Phi_{\gamma_b+\gamma_c}(\nu_{J_b \mu_{b}J_a\mu_a}-\xi^\prime)
+\Phi^{\ast}_{\gamma_{b}+\gamma_c}(\nu_{J_b \mu_{b^\prime}J_a\mu_a}-\xi^\prime)\right].
\label{qed_gen_prof-mstate}
\end{eqnarray}
The $\Phi^{K,K^{\prime}}_{Q}$ defined above can be seen as a frequency-dependent 
coupling coefficient that connects the $(K^\prime,Q)$ multipole component 
of the incident radiation field with the $(K,Q)$ multipole moment of 
the atomic density matrix (see LL04, p. 525). In the non-magnetic and 
weak field limits, the $\nu_L$ dependence of the profile function 
$\Phi_{\gamma_b+\gamma_c}$ can be neglected, which gives us 
\begin{equation}
{\rm lim}_{\nu_L \to 0} \Phi^{K,K^\prime}_Q(a,b;\xi') = 
\delta_{KK^\prime} w^{(K)}_{ba} \phi(\nu_0-\xi'),
\label{gen_prof_weaklimit}
\end{equation}
where $w^{(K)}_{ba}$ is defined in Equation~(10.11) of LL04, and 
$\phi$ denotes the usual non-magnetic profile function. In this
limit we have $K^\prime=K$.

In the case of $J$-state interference a suitable quantum generalized 
profile function has not been defined yet, but we can define it here in
analogy with the two-level atom case.
It has the following form for the incoming radiation:
\begin{eqnarray}
&&\Phi^{K,K^\prime}_Q(a,{b^\prime},b;\xi^\prime) = (2J_a+1)
\sqrt{3(2J_{b^\prime}+1)(2J_b+1)(2K+1)(2K^\prime+1)}
\sum_{\mu_b\mu_{b^\prime} \mu_a q'q'''} (-1)^{1+J_b-\mu_{b^\prime}+q'}
\nonumber \\ &&\times
\left(
\begin{array}{ccc}
J_b & J_a & 1 \\
-\mu_b & \mu_a & -q' \\
\end{array}
\right)
\left(
\begin{array}{ccc}
J_{b^\prime} & J_a & 1 \\
-\mu_{b^\prime} & \mu_a & -q''' \\
\end{array}
\right)
\left(
\begin{array}{ccc}
J_b & J_{b^\prime} & K \\
\mu_b & -\mu_{b^\prime} & -Q \\
\end{array}
\right)
\left(
\begin{array}{ccc}
1 & 1 & K^\prime \\
q''' & -q' & -Q \\
\end{array}
\right)\nonumber \\ &&\times
{1\over 2} \left[\Phi_{\gamma_b+\gamma_c}
(\nu_{b_ma_m}-\xi^\prime)
+\Phi^{\ast}_{\gamma_{b^\prime}+\gamma_c}
(\nu_{{b^\prime}_ma_m}-\xi^\prime)\right]
\cos\beta_{{b'}_mb_m} {\rm e}^{{\rm i}\beta_{{b'}_mb_m}},
\label{qed_gen_prof-jstate}
\end{eqnarray}
with a similar expression for the outgoing radiation when $J_a$ and $\mu_a$ 
are replaced respectively by $J_f$ and $\mu_f$, and angle 
$\beta_{{b'}_mb_m}$ is replaced by $\alpha_{{b'}_mb_m}$. 
Notice that unlike the two-level atom case we now have included the angles 
$\beta_{{b'}_mb_m}$ and $\alpha_{{b'}_m{b}_m}$ in 
the definition of the quantum generalized profile function, as they cannot 
be taken outside the summation over the magnetic substates. 
Using the orthogonality 
relation of the $3-j$ symbols, it is easy to verify that 
\begin{eqnarray}
&&\sum_K\Phi^{K,K^\prime}_Q(a,{b^\prime},b;\xi^\prime)
\Phi^{K,K^{\prime\prime}}_Q(f,{b^\prime},b;\xi)
= 3G\  \sqrt{(2K^\prime+1)(2K^{\prime\prime}+1)}\nonumber \\ && \times
\sum_{\mu_b\mu_{b^\prime} \mu_a \mu_f qq^\prime q^{\prime\prime}
q^{\prime\prime\prime}} Z_3\ \ 
(-1)^{q^{\prime}+q^{\prime\prime}+Q}
\left(
\begin{array}{ccc}
1 & 1 & K^{\prime\prime} \\
q & -q^{\prime\prime} & Q \\
\end{array}
\right)
\left(
\begin{array}{ccc}
1 & 1 & K^{\prime} \\
q^{\prime\prime\prime} & -q^{\prime} & -Q \\
\end{array}
\right) \nonumber \\ &&\times
{1\over 2} \bigg[\Phi_{\gamma_b+\gamma_c}
(\nu_{b_ma_m}-\xi^\prime)
+\Phi^{\ast}_{\gamma_{b^\prime}+\gamma_c}
(\nu_{{b^\prime}_ma_m}-\xi^\prime)\bigg]
{1\over 2} \bigg[\Phi_{\gamma_b+\gamma_c}(\nu_{b_mf_m}-\xi)
\nonumber \\ && +\Phi^\ast_{\gamma_{b^\prime}+\gamma_c}
(\nu_{{b^\prime}_mf_m}-\xi)\bigg] 
\cos\beta_{{b'}_mb_m}\cos\alpha_{{b'}_mb_m}
{\rm e}^{{\rm i}(\beta_{{b'}_mb_m}+\alpha_{{b'}_mb_m})},
\label{qed_gen_prof-jstate-prod}
\end{eqnarray}
which is a useful relation that helps in the identification of the multipolar 
index $K$.
An equivalent relation, but for the case of $m$-state interference, is 
Equation~(22) of \citet{vb97b}. 

{Comparing the terms in the flower brackets of Equation~(\ref{pehn_r3-1}) with the 
RHS of Equation~(\ref{qed_gen_prof-jstate-prod}), we can see that they are the
same (except for some factors). Therefore after substituting the 
terms in the flower brackets of Equation~(\ref{pehn_r3-1}) with the LHS of 
Equation~(\ref{qed_gen_prof-jstate-prod}),
we assume $\gamma_c/2=D^{(K)}$ as a reasonable approximation (see S94),
where $D^{(K)}$ is the $2K$ multipole collisional 
destruction rate. Further, following BS99,
we identify $\gamma_b = \Gamma_{R{b}}+\Gamma_{I{b}}$
and $\gamma_{c}=\Gamma_{E}$, where $\Gamma_{Rb}$ is the radiative
width of the fine structure state $J_b$.
$\Gamma_{Ib}$ is the total inelastic collision rate
defined for the state $J_b$. It is given by
\begin{equation}
\Gamma_{Ib}=\sum_{f}\Gamma_{I{bf}} + 
\sum_{{b^\prime} \ne b} \Gamma_{I{b{b^\prime}}}.
\label{gammai-jb}
\end{equation}
Here $\Gamma_{I{bf}}$ couple the upper state $J_b$ to the lower state $J_f$ and 
$\Gamma_{I{b{b^\prime}}}$ couples the two fine structure states $J_b$ and 
$J_{b^\prime}$. 
Indeed such a definition of total inelastic collision rates can be found in
\citet{osc72} and also in Equations~(2.15)-(2.20) of \citet{hh82}.
$\Gamma_{E}$ is the elastic collision rate and $D^{(K)}$ represent the 
depolarizing elastic collisions that couple the Zeeman substates 
($m$-states) of a given $J_b$-state.  
In general $D^{(K)}$ may be different for each of the fine structure components 
with
quantum number $J_{b}$. {However, as an approximation we assume them to be
independent of the $J$-quantum number.}

With the above mentioned substitutions and identifications,
Equation~(\ref{pehn_r3-1}) can be rewritten as 
\begin{eqnarray}
&&{\bf R}^{\rm III}_{ij}(\xi,{\bm n}; \xi^\prime,{\bm n}^\prime,{\bm B})=
{ 2\over 9}(2L_a+1)^{2}
\sum_{KK^\prime K^{\prime\prime}Qafb{b^\prime}}
B^{(K)}_{b{b^\prime}} Z_6
(-1)^Q{\mathcal T}^{K^{\prime\prime}}_Q(i,{\bm n})
{\mathcal T}^{K^{\prime}}_{-Q}(j,{\bm n}^\prime) \nonumber \\ && \times
\Phi^{K,K^\prime}_Q(a,{b^\prime},b;\xi^\prime)
\Phi^{K,K^{\prime\prime}}_Q(f,{b^\prime},b;\xi),\ \ \ \ \ \ \ 
\label{pehn_r3_k}
\end{eqnarray}
where $B^{(K)}_{b{b^\prime}}$ is the collisional branching ratio 
defined as 
\begin{equation}
B^{(K)}_{b{b^\prime}}=
\frac{\overline{\Gamma}_{R}^{b{b^\prime}}}
{\overline{\Gamma}_{R}^{b{b^\prime}}
+\overline{\Gamma}_{I}^{b{b^\prime}}+D^{(K)}}\frac{\Gamma_{E}-D^{(
K)}}{\overline{\Gamma}_{R}^{b{b^\prime}}+\overline{\Gamma}_{I}^{b{b^\prime}}
+\Gamma_{E}}.
\label{branch-b}
\end{equation}
Also, the branching ratio $A_{b{b^\prime}}$ can be written as
\begin{equation}
 A_{b{b^\prime}}=\frac{\overline{\Gamma}_{R}^{b{b^\prime}}}
{\overline{\Gamma}_{R}^{b{b^\prime}} + \overline{\Gamma}_{I}^{b{b^\prime}} 
+ \Gamma_E},
\label{branch-a}
\end{equation}
where
\begin{equation}
\overline{\Gamma}_{R}^{b{b^\prime}}=\frac{\Gamma_{Rb} + \Gamma_{R{b^\prime}}}{2} ; 
\quad \overline{\Gamma}_{I}^{b{b^\prime}}=\frac{\Gamma_{Ib} + 
\Gamma_{I{b^\prime}}}{2},
\label{ave-gam}
\end{equation}
with $\Gamma_{Ib}$ defined in Equation~({\ref{gammai-jb}).
The total damping rates that appear in the branching ratios
 are the same as those
that appear in the denominator of the Hanle angles
 (see Equations~(\ref{angle_alpha-beta}) and (\ref{alpha-k})).
Therefore the $J_bJ_{b^\prime}$ dependence of the branching ratios defined 
now for a 
two-term atom is self-consistent.
We have verified that when we set $J_b=J_{b^\prime}$ and 
$J_a=J_f$ (the case of a two-level atom with only $m$-state interference) 
in Equation~(\ref{pehn_r3_k}), we recover Equation~(49) of \citet{vb97b}.}}

\subsection{The laboratory frame expression for the  
redistribution matrix ${\bf R}^{\rm III}$}
\label{lab-redis}
We convert Equation~(\ref{pehn_r3_k}) into the laboratory frame using 
the same procedure as described in Section~2.2 of HZ2. The resulting 
expression for the normalized type-III redistribution matrix in the 
laboratory frame can be written as
\begin{eqnarray}
&&{\bf R}^{\rm III}_{ij}( x,{\bm n}; x^\prime,{\bm n}^\prime,{\bm B})=
{ 2L_b+1\over 2S+1}
\sum_{KK^\prime K^{\prime\prime}Qafb{b^\prime}} Z_6\ \ 
B^{(K)}_{b{b^\prime}}
(-1)^Q{\mathcal T}^{K^{\prime\prime}}_Q(i,{\bm n})
{\mathcal T}^{K^{\prime}}_{-Q}(j,{\bm n}^\prime) \nonumber \\ && \times
{\mathcal {R}}^{K^{\prime \prime},K,K^{\prime}}_{Q,{\rm III}}(x,x^{\prime},\Theta,{\bm B}),
\label{r3_lab}
\end{eqnarray}
where ${\mathcal {R}}^{K^{\prime \prime},K,K^{\prime}}_{Q,{\rm III}}(x,x^{\prime},
\Theta,{\bm B})$ is the laboratory frame redistribution function obtained after 
transformation of the atomic frame functions 
$\Phi^{K,K^\prime}_Q(a,{b'},b;\xi^\prime)
\Phi^{K,K^{\prime\prime}}_Q(f,{b'},b;\xi)$. The factors $(2L_b+1)/(2S+1)$ 
result from the renormalization of Equation~(\ref{pehn_r3_k}). The function 
${\mathcal {R}}^{K^{\prime \prime},K,K^{\prime}}_{Q,{\rm III}}$ has the 
following form\,:
\begin{eqnarray}
&& {\mathcal {R}}^{K^{\prime \prime},K,K^{\prime}}_{Q,{\rm III}}
(x,x^{\prime},\Theta,{\bm B})= \sum_{\mu_a\mu_f\mu_b\mu_{b^{\prime}}\mu_{b^{\prime\prime}}
\mu_{b^{\prime\prime\prime}}qq^{\prime}q^{\prime\prime}q^{\prime\prime\prime}}
{3\over 4} G 
(2K+1)
 \sqrt{(2K^\prime+1)(2K^{\prime\prime}+1)} (-1)^{1+J_b-\mu_{b^\prime}+q'} 
\nonumber \\ && \times
(-1)^{1+J_b-\mu_{b^{\prime\prime\prime}}+q}
\cos\beta_{{b'}_mb_m}\cos\alpha^{(K)}_
{{b^{\prime\prime\prime}_m}
{b^{\prime\prime}_m}}
{\rm e}^{{\rm i}\left(\beta_{{b'}_mb_m}+
\alpha^{(K)}_{{b^{\prime\prime\prime}_m}{b^{\prime\prime}_m}}\right)}  
\nonumber \\ && \times
\left(
\begin{array}{ccc}
J_b & J_a & 1 \\
-\mu_b & \mu_a& -q' \\
\end{array}
\right)
\left(
\begin{array}{ccc}
J_{b'} & J_a & 1 \\
-\mu_{b'} & \mu_a & -q''' \\
\end{array}
\right)
\left(
\begin{array}{ccc}
J_b & J_{b^\prime} & K \\
\mu_b & -\mu_{b^\prime} & -Q \\
\end{array}
\right)
\left(
\begin{array}{ccc}
1 & 1 & K^\prime \\
q''' & -q' & -Q \\
\end{array}
\right) \nonumber \\ && \times
\left(
\begin{array}{ccc}
J_{b} & J_f & 1 \\
-\mu_{b^{\prime\prime}} & \mu_f & -q \\
\end{array}
\right)
\left(
\begin{array}{ccc}
J_{b^\prime} & J_f & 1 \\
-\mu_{b^{\prime\prime\prime}} & \mu_f & -q'' \\
\end{array}
\right) 
\left(
\begin{array}{ccc}
J_b & J_{b'} & K \\
\mu_{b^{\prime\prime}} & -\mu_{b^{\prime\prime\prime}} & -Q \\
\end{array}
\right)
\left(
\begin{array}{ccc}
1 & 1 & K^{\prime\prime} \\
q^{\prime\prime} & -q & -Q \\
\end{array}
\right)
\nonumber \\ && \times
\Bigg\{\left[
R^{\rm III,\,HH}_{{b^\prime}_ma_m,b^{\prime\prime\prime}_mf_m}+
R^{\rm III,\,HH}_{{b^\prime}_ma_m,{b^{\prime\prime}_mf_m}}+
R^{\rm III,\,HH}_{b_ma_m,{b^{\prime\prime\prime}_mf_m}}+
R^{\rm III,\,HH}_{b_ma_m,{b^{\prime\prime}_m}f_m}\right]\nonumber \\ && 
+{\rm i}\left[
R^{\rm III,\,FH}_{{b^\prime}_ma_m,b^{\prime\prime\prime}_mf_m }+
R^{\rm III,\,FH}_{{b^\prime}_ma_m,{b^{\prime\prime}_m}f_m}-
R^{\rm III,\,FH}_{b_ma_m,{b^{\prime\prime\prime}_m}f_m}-
R^{\rm III,\,FH}_{b_ma_m,{b^{\prime\prime}_m}f_m}\right]\nonumber \\ &&
+{\rm i}\left[R^{\rm III,\,HF}_{{b^\prime}_ma_m,
{b^{\prime\prime\prime}_m}f_m}-
R^{\rm III,\,HF}_{{b^\prime}_ma_m,
{b^{\prime\prime}_m}f_m}+
R^{\rm III,\,HF}_{b_ma_m,{b^{\prime\prime\prime}_m}f_m}-
R^{\rm III,\,HF}_{b_ma_m,{b^{\prime\prime}_m}f_m}\right]\nonumber \\ &&
-\left[R^{\rm III,\,FF}_{{b^\prime}_ma_m,
b^{\prime\prime\prime}_mf_m}-
R^{\rm III,\,FF}_{{b^\prime}_ma_m,
{b^{\prime\prime}_m}f_m} -
R^{\rm III,\,FF}_{b_ma_m,{b^{\prime\prime\prime}_m}f_m}+
R^{\rm III,\,FF}_{b_ma_m,{b^{\prime\prime}_m}f_m}\right]\Bigg\}.
\end{eqnarray}
The results presented in Section~{\ref{results}} are computed using 
Equation~(\ref{r3_lab}) of the present paper for the type-III redistribution and 
Equation~(25) of P1 for the type-II redistribution matrix. The form of the total 
redistribution matrix in the laboratory frame remains the same as 
Equation~(\ref{phen_rm}).
In the non-magnetic case Equation~(\ref{r3_lab}) simplifies to 
\begin{eqnarray}
&&{\bf R}^{\rm III}_{ij}(x,{\bm n};x^\prime, {\bm n}^\prime) = 
\frac{3(2L_b+1)}{2S+1}
\sum_{KQafb{b^\prime}}\  G\  Z_6\ \ (-1)^{J_f-J_a} 
 B^{(K)}_{b{b^\prime}} \cos\beta_{{b'}b} 
\cos\alpha^{(K)}_{{b'}b} 
{\rm e}^{{\rm i}\left(\beta_{{b'}b}+\alpha^{(K)}_{{b'}b}\right)} \nonumber
 \\ && \times
\Big[h^{\rm III}_{ba,{b^\prime}f}
+ {\rm i} f^{\rm III}_{ba,{b^\prime}f}
\Big]
\left\{
\begin{array}{ccc}
1 & 1 & K \\
J_{b^\prime} & J_b & J_a \\
\end{array}
\right\}
\left\{
\begin{array}{ccc}
1 & 1 & K \\
J_{b^\prime} & J_b & J_f \\
\end{array}
\right\} (-1)^{Q} {\mathcal T}^K_Q(i,{\bm n})
{\mathcal T}^{K}_{-Q}(j,{\bm n}^\prime).
\label{dh_mat}
\end{eqnarray}
The angle $\beta_{{b'}b}$ and the auxiliary functions
 $h^{\rm III}_{ba,{b'}f}$ and $f^{\rm III}_{ba,{b'}f}$ 
are defined respectively in Equations~(\ref{angle_alpha-beta}), (\ref{h3mubmubprimemuf}) and 
(\ref{f3mubmubprimemuf}), but with $\nu_L=0$.
{The angles $\alpha^{(K)}_{{b^\prime}b}$ and 
$\beta_{{b^\prime}b}$ (arising exclusively
from $J$-state interference) are defined as
\begin{equation}
\tan \alpha^{(K)}_{{b^\prime}b}= \frac{\omega_{{b^\prime}b}} 
{\overline{\Gamma}_{R}^{{b^\prime}{b}}+\overline{\Gamma}_{I}^{{b^\prime}b}+D^{(K)}};
\quad \tan \beta_{{b^\prime}b}= \frac{\omega_{{b^\prime}b}} 
{\overline{\Gamma}_{R}^{{b^\prime}{b}}+\overline{\Gamma}_{I}^{{b^\prime}b}+\Gamma_{E} },
\label{alpha-k}
\end{equation}
with $\overline{\Gamma}_{R}^{{b^\prime}{b}}=\overline{\Gamma}_{R}^{b{b^\prime}}$
and $\overline{\Gamma}_{I}^{{b^\prime}{b}}=\overline{\Gamma}_{I}^{b{b^\prime}}$
(see Equation~(\ref{ave-gam})).
In P3, the two-level atom branching ratios were used in the realistic 
modeling of the linear polarization profiles of the Cr\,{\sc i} triplet. 
These branching ratios can be recovered 
from the more general two-term atom expressions given in
Equations~(\ref{branch-b}) and (\ref{branch-a})
by neglecting $\Gamma_{Ib{b^\prime}}$. This is equivalent
to setting $J_b=J_{b^\prime}$
in  Equations~(\ref{branch-b}) and (\ref{branch-a}).}
{The angle-averaged redistribution matrices corresponding
to the angle-dependent redistribution matrices presented
in Equations~(\ref{r3_lab})-(\ref{dh_mat}) can be recovered
by replacing the angle-dependent redistribution functions
(Equations~(\ref{rhh3})-(\ref{rff3}))
by their angle-averaged analogues.}
These angle-averaged functions are obtained
by numerical integration of the angle-dependent
functions over the scattering angle $\Theta$ (see Equation~(\ref{aa-rf})). 

\section{The polarized radiative transfer equation for $J$-state interference}
\label{rt-gen}
 The polarized radiative transfer equation for the Stokes vector
${\bm I}$ in a one-dimensional planar medium for the Hanle scattering problem 
can be written as
\begin{equation}
 \mu\frac{\partial{\bm I}(\tau,x,{\bm n})}{\partial \tau}= (\phi(x) + 
{r})[{\bm I}(\tau, x, {\bm n})-{\bm S}(\tau, x,{\bm n})],
\label{rte-1}
\end{equation}
where the notations are the same as those used in P2, with the
positive Stokes $Q$ representing electric vector vibrations perpendicular 
to the solar limb. {This definition is opposite to the way in which the positive
Stokes $Q$ is defined in the observed spectra. This can easily be accounted
for (through a sign change), when comparing the observed spectra with the theoretical
results.} ${\bm n}=(\vartheta,\varphi)$ defines the ray direction
where $\vartheta$ and $\varphi$ are the inclination
and azimuth of the scattered ray with $\mu={\rm cos}\,\vartheta$ (see Figure~1 of P1).
In the weak magnetic field limit, the Stokes vector ${\bm I}=(I,Q,U)^{\rm T}$ and the 
Stokes
source vector ${\bm S}=(S_I,S_Q,S_U)^{\rm T}$. 
In this limit, the transfer equation for Stokes $V$ decouples from that of
the Stokes vector $(I,Q,U)^{\rm T}$. 
This is known as the weak field approximation. 
In Equation~(\ref{rte-1}), the Stokes vector ${\bm I}$
and the Stokes source vector ${\bm S}$ depend on ${\bm n}$. 
In the case of angle-averaged redistribution, it was shown by \citet{hf07}
that one can decompose ${\bm S}$ and ${\bm I}$ into six cylindrically symmetric
components $\mathcal{I}^K_Q$ and $\mathcal{S}^K_Q$ with the help 
of the irreducible spherical tensors for polarimetry
(See \citet{landi84}). Here, $K=0,2$ and $-K\le Q \le +K$.
Such a decomposition results in a reduced Stokes vector  $\bm {\mathcal{I}}$
which is independent of $\varphi$ and a reduced source
vector $\bm {\mathcal{S}}$ which is independent of both $\vartheta$ and $\varphi$. 
{We denote the quantities in the reduced basis by calligraphic letters and
in Stokes basis by Roman.}
In such a reduced basis the transfer equation can be written as
\begin{equation}
  \mu\frac{\partial{\bm {\mathcal I}}(\tau,x,\mu)}{\partial \tau}= (\phi(x) + 
{r})[{\bm{\mathcal I}}(\tau, x, \mu)-{\bm {\mathcal S}}(\tau, x)].
\label{tkq-rte}
\end{equation}
The reduced source vector is defined as
\begin{equation}
 {\bm {\mathcal S}}(\tau,x)=\frac{\phi(x){\bm{\mathcal S_{l}}(\tau,x)}+ 
{r}{\bm{\mathcal G}(\tau)}}{\phi(x)+{r}},
\end{equation}
where ${\bm {\mathcal G}}(\tau)=\{B,0,0,0,0,0\}^{\rm T}$ is the primary source 
vector.
The {reduced} line source vector is given by
\begin{equation}
{\bm{\mathcal S_{l}}}(\tau,x)=\sum_{b{b^\prime}}
\bigg[\epsilon_{b{b^\prime}}{\bm{\mathcal G}}(\tau) + \int^{+\infty}_{-\infty}
\frac{{{\bm {\mathcal R}}}_{b{b^\prime}}(x,x^{\prime},{\bm B})}{\phi(x)}
{\bm {\mathcal J}}(\tau,x^{\prime})dx^{\prime}\bigg], 
\label{source-l}
\end{equation}
{ where ${{\bm {\mathcal R}}}_{b{b^\prime}}(x,x^{\prime},{\bm B})$
is the redistribution matrix for a two-term atom, with
the summation over $J_b$ and $J_{b^\prime}$ not yet performed.
The thermalization parameter is given by
\begin{equation}
 \epsilon_{b{b^\prime}}=\frac{\overline{\Gamma}_{I}^{b{b^\prime}}}
{\overline{\Gamma}_{R}^{b{b^\prime}} +  \overline{\Gamma}_{I}^{b{b^\prime}} }.
\label{eps}
\end{equation}
 The computation of the above defined reduced line source vector is 
very expensive because of the summations over $J_b$ and $J_{b^\prime}$
which need to be performed at each iteration.
However for all practical applications, we can assume $\epsilon$ to be the same 
for all the $J_bJ_{b^\prime}$ states, which is a good approximation.
 Such an approximate $\epsilon$ 
is constructed by taking an average value of $\Gamma_{Ib}$ for all
transitions involving $J_b,J_{b^\prime}$ and $J_f$ 
and an average value of $\Gamma_R$ for all the upper fine structure states.
Under such an approximation, the reduced line source vector can 
be written as} 
\begin{equation}
{\bm{\mathcal S_{l}}}(\tau,x)=\epsilon{\bm{\mathcal G}}(\tau) + \int^{+\infty}_{-\infty}
\frac{{{\bm {\mathcal R}}}(x,x^{\prime},{\bm B})}{\phi(x)}
{\bm {\mathcal J}}(\tau,x^{\prime})dx^{\prime}.
\label{source-l-1}
\end{equation}  
The mean intensity $\bm{\mathcal J}(\tau, x)$ 
is defined by
\begin{equation}
 {\bm{\mathcal J}}(\tau,x)=\frac{1}{2}\int^{+1}_{-1}{\bf \Psi}(\mu^{\prime})
 {\bm{\mathcal I}}(\tau,x,\mu^{\prime})d \mu^{\prime}.
\end{equation}
The elements of the ${\bf \Psi(\mu)}$ matrix are given in LL04
(see also Appendix A of \citet{hf07}).
${{\bm {\mathcal R}}}(x,x^{\prime}, {\bm B})$ appearing in 
Equation~(\ref{source-l}) 
is a $(6\times 6)$ diagonal matrix.
The explicit form of this redistribution matrix with and without the presence of 
magnetic fields {is defined} in the following sections.
In the absence of a magnetic field, only the ${\mathcal{I}^{0}_{0}}$ and
${\mathcal{I}^{2}_{0}}$
components contribute to the Stokes vector. Hence the $(6 \times 6)$ problem
reduces to a $(2 \times 2)$ problem. 
The transfer equation defined in Equation~({\ref{tkq-rte}}) is solved 
using the traditional polarized accelerated lambda iteration technique
presented in P2.

\subsection{The redistribution matrix ${{\bm {\mathcal R}}}(x,x^{\prime})$ 
for the non-magnetic case}
\label{rt-nonmag}
In the absence of a magnetic field the redistribution matrix in Equation~(\ref{source-l}) 
becomes independent of ${\bm B}$ and reduces to a $(2\times 2)$ diagonal matrix
with elements  ${{\bm {\mathcal R}}}(x,x^{\prime})$=diag
$({\mathcal R}^{0},{\mathcal R}^{2})$. The elements ${\mathcal R}^{K}$
are defined as
\begin{eqnarray}
&& {\bf {\mathcal R}}^{K}(x,x^\prime) = \frac{3(2L_b+1)}{2S+1}
\sum_{b{b^\prime}af} G Z_6 (-1)^{J_f-J_a} 
\left\{
\begin{array}{ccc}
1 & 1 & K \\
J_{b^\prime} & J_b & J_a \\
\end{array}
\right\}
\left\{
\begin{array}{ccc}
1 & 1 & K \\
J_{b^\prime} & J_b & J_f \\
\end{array}
\right\}\nonumber \\ && \times
\Big\{A_{b{b^\prime}} \ 
\cos\beta_{{b^\prime}b} \ \ 
{\rm e}^{{\rm i}\beta_{{b^\prime}b}}
\Big[(h^{\rm II}_{b,{b^\prime}})_{af}
+{\rm i}(f^{\rm II}_{b,{b^\prime}})_{af}
\Big]\ \nonumber \\ &&
+ B^{(\rm K)}_{b{b^\prime}} \cos\beta_{{b^\prime}b}\ 
\cos\alpha^{(K)}_{{b^\prime}b}\ \ 
{\rm e}^{{\rm i}(\beta_{{b^\prime}b}+
\alpha^{(K)}_{{b^\prime}b})} 
\Big[h^{\rm III}_{ba,{b^\prime}f}
+ {\rm i} f^{\rm III}_{ba,{b^\prime}f}
\Big]\Big\} .
\label{r_mat}
\end{eqnarray}
The $(h^{\rm II}_{b,{b^\prime}})_{af}$ and 
$(f^{\rm II}_{b,{b^\prime}})_{af}$ are the auxiliary functions 
for type-II defined in
Equations~(14) and (15) of P1 and the auxiliary functions for type-III are
defined in Equations~(\ref{h3mubmubprimemuf})
and (\ref{f3mubmubprimemuf}) of the present paper. They are used here for the 
non-magnetic case
and with {the} angle-averaged redistribution functions of type-II and type-III.
In the limit of a two-level atom model 
($J_b=J_b'$ and $J_a=J_f$), the $(h^{\rm II}_{b,{b^\prime}})_{af}$
and $h^{\rm III}_{ba,{b^\prime}f}$ go respectively
to $R^{\rm II}$ and $R^{\rm III}$ functions of Hummer,
whereas the $(f^{\rm II}_{b,{b^\prime}})_{af}$ 
and $f^{\rm III}_{ba,{b^\prime}f}$ and the angles
$\beta_{{b^\prime}b}$ and $\alpha^{(K)}_{{b^\prime}b}$
go to zero.

\subsection{The redistribution matrix 
${{\bm {\mathcal R}}}(x,x^{\prime}, {\bm B})$ 
for the magnetic case}
\label{rt-mag}
\begin{figure}
\centering
\includegraphics[width=10.0cm,height=6.0cm]{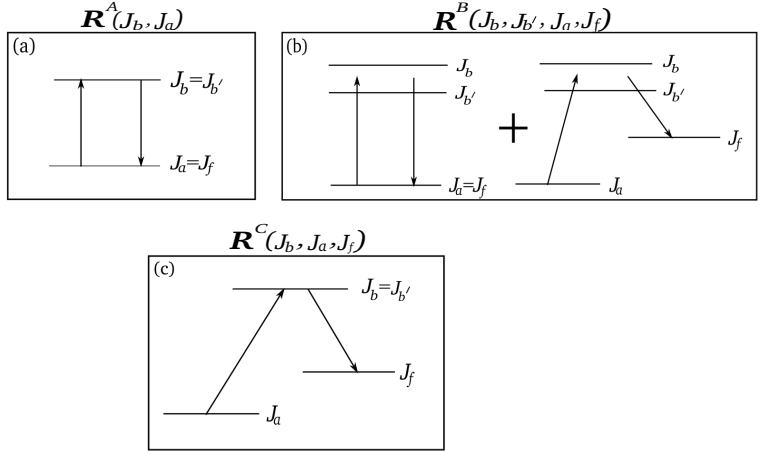}
\caption{The {schematic} level diagrams representing the three components of the 
$J$-state redistribution matrix. Panels (a), (b), and (c) represent
respectively the two-level atom resonance scattering, $J$-state 
interference, and Raman scattering (resonance fluorescence).}
\label{level-diag-redis}
\end{figure}

{The redistribution matrix for a two-term atom defined in
Equation~(\ref{phen_rm}) 
involves summations over the total angular momentum quantum numbers 
and the corresponding 
magnetic 
quantum numbers.} This does not allow direct decomposition to
go from {the} Stokes vector basis to the reduced basis
(see Section~{\ref{rt-gen}} for details on these two basis).
Such a decomposition is possible in the non-magnetic case.
This is because in the absence of a magnetic field, the summations
over the magnetic quantum numbers can be analytically
performed using Racah algebra as shown in P1 for type-II redistribution
and Equation~(\ref{dh_mat}) for type-III redistribution.
However in the magnetic case, all the summations remain intact
and have to be performed numerically. This is very expensive.
Because of these difficulties, we need to resort to the 
weak field approximation which allows us to apply {the} decomposition technique. 
In this regard, the summations over the $J$-quantum numbers
can be split into three different terms namely
\begin{eqnarray}
{\bm {\mathcal{R}}}(x,x^\prime,\bm B)=\!\!\!\!\!\!
\sum_{b={b^\prime},a=f}
\bm{\mathcal{R}}^{\rm A}_{b,a}(x,x',\bm B) + \!\!\!\!\!\!
\sum_{b\ne {b^\prime},a,f} 
\bm{\mathcal{R}}^{\rm B}_{b,{b^\prime},a,f}(x,x',\bm B) + \!\!\!\!\!\!
\sum_{b={b^\prime},a\ne f}
\bm{\mathcal{R}}^{\rm C}_{b,a,f}(x,x',\bm B).
\label{redis-tot}
\end{eqnarray}
The first term represents the case of `Resonance' scattering
in a two-level atom model with a summation over 
all the lines of the multiplet (see Figure~{\ref{level-diag-redis}a}). 
This contributes mainly to the cores and near wings of the lines within the multiplet.
Its weak field analogue has already been derived in \citet{vb97b} and can be 
expressed as
\begin{eqnarray}
{\bm{\mathcal{R}}}^{\rm A}(x,x^\prime,\bm B)=\sum_{ab}\bm{\mathcal{R}}^{\rm A}_{b,a}(x,x',\bm B)= 
\sum_{ab} \mathcal{W}_{b,a} \bm{\mathcal{R}}^{\rm  H}_{b,a}(x,x',\bm B).
\label{redis-two-level}
\end{eqnarray}
Here $\bm{\mathcal{R}}^{\rm  H}_{b,a}(x,x',\bm B)$
is the Hanle redistribution matrix
for a two-level atom with $J_a \to J_b \to J_a$ 
scattering transition as presented in \citet{vb97b}.
This is also the same redistribution matrix defined in Appendix A of \citet{lsa11},
 but for a $J_a \to J_b\to J_a$
scattering transition. The details of the domain based decomposition of this
matrix are also given in the above paper.
In the reduced basis, ${\bm{\mathcal{R}}}^{\rm A}(x,x^\prime,\bm B)$
is a $(6 \times 6)$ matrix.
$\mathcal{W}_{b,a}$
are the weights for each line component of the multiplet 
(derived from Equation~(\ref{r_mat}) with $J_b=J_{b^\prime}$ and $J_a=J_f$)
and are given by
\begin{eqnarray}
\mathcal{W}_{b,a} &=& \frac{(2L_b+1)}{2S+1}
(2J_a+1)^2(2J_b+1)^2 \left\lbrace 
\begin{array}{ccc}
L_a & L_b & 1\\
J_b & J_a & S \\
\end{array}
\right\rbrace^4.
\end{eqnarray}

The second term represents only the $J$-state interference 
between different lines of the multiplet (see Figure~{\ref{level-diag-redis}b}).
It includes both {the} `Resonance' and {the} `Raman' scattering parts and
is effective mainly in the wings between the lines. 
This term is quite insensitive
to the strength of the magnetic field. 
This can be seen from Figures~2 and 3 of P1 where
the magnetic field effects are confined mainly to the line cores. 
Hence in this component we can set the magnetic field equal to zero
as a good approximation.
This makes the evaluation of both {the} Resonance and {the} Raman scattering parts
similar to that performed in P2 and P3.
In the reduced basis this term 
is simply given by 
a $(6 \times 6)$ matrix ${{\bm {\mathcal R}}^{\rm B}(x,x^{\prime})}$  
which is equal to $\rm{diag} ({\mathcal R}^{0},{\mathcal R}^{2},0,0,0,0)$.
Here ${\mathcal R}^{K}$ are the redistribution functions which include the effects 
of collisions and {the} $J$-state interference between different line 
components in a multiplet defined in Equation~(\ref{r_mat}). Only $J_b \ne J_b^\prime$
terms are retained in the summations appearing in this equation. The $J_b=J_{b^\prime}$
contributions are contained in the first term. 
However in some of the well known examples in the second solar spectrum
like the Mg\,{\sc ii} h and k, Ca\,{\sc ii} H and K and the Cr\,{\sc i} triplet, 
the initial and the final states are the same. Also for the case of the 
hypothetical doublet
considered in this paper, arising due to an $L=0 \to 1 \to 0$ scattering transition
with spin $S=1/2$, the initial and
the final states are the same. Hence the Raman scattering
part does not play a role.

The third term represents the case of only Raman scattering without {the} $J$-state
interference, where the initial and the final states are different 
(see Figure~{\ref{level-diag-redis}c}).
A derivation of the weak field analogue of this component
(in a way similar to that of \citet{vb97b}) is yet to be performed. 
Again for some of the well known examples mentioned above, this component
 does not contribute.
Thus the final expression for the redistribution matrix that is used in 
Equation~(\ref{source-l})
is
\begin{equation}
{\bm{\mathcal{R}}}(x,x^\prime,\bm B) \approx 
{\bm{\mathcal{R}}}^{\rm A}(x,x^\prime,\bm B)+
{\bm{\mathcal{R}}}^{\rm B}(x,x^\prime).
\end{equation}

\section{Results and discussion}
\label{results}
In this section, we study the effects of collisional
redistribution matrix on the emergent Stokes profiles for the case of single scattering 
and also multiple scattering in an isothermal atmospheric slab.
All the profiles presented in this paper are computed for a
hypothetical doublet line system with {the} line center wavelengths
at 5000\,\AA\ and 5001\,\AA\ arising due to an $L=0 \to 1 \to 0$
scattering transition with spin $S=1/2$. The $J$ quantum numbers of the
 lower and upper states are
$J_{a}=J_{f}=1/2$ and $J_{b}=1/2, 3/2$. In Section~{\ref{sing-scat}}
we present the scattered Stokes profiles resulting in a single $90^\circ$ scattering
{ case}. In Section~{\ref{iso-rt}} we present the
multiply scattered Stokes profiles emerging from an isothermal constant property
atmospheric slab with and without the presence of a magnetic field. 

\subsection{The single $90^{\circ}$ scattering case}
\label{sing-scat}
To explore the general behavior of the redistribution matrix in the presence of
collisions we illustrate the Stokes profiles that result from single $90^{\circ}$ 
scattering event. We examine the influence of the elastic collisions 
on the Stokes profiles in the presence of a magnetic field. The magnetic
 field orientation is given by 
$\vartheta_{B}=90^{\circ}$ and $\varphi_{B}=45^{\circ}$ where the  colatitude
$\vartheta_{B}$ and azimuth $\varphi_{B}$ characterize the magnetic 
field orientation with respect to the polar 
$z$-axis (see Figure~1 of P1).
We consider an unpolarized (${\bm I_{in}}=[1,0,0,0]^{\rm T}$) 
and spectrally flat (frequency independent) radiation field that is incident
 in the vertical
direction (parallel to the polar $z$-axis). The singly scattered
Stokes vectors are then exclusively determined by the first column of
the angle-dependent redistribution matrix {by integrating over the 
incident wavelengths}.
However in the multiple scattered solutions discussed in { Section~{\ref{iso-rt}},
we} restrict our attention only to the angle-averaged redistribution matrix.
The magnetic field strength is parametrized by the splitting parameter 
$v_{\rm H}$ given by 
\begin{equation}
 v_{\rm H}=\frac{\lambda_{0}^{2}e_{0}B}{4\pi m c^{2}}\times
\frac{1}{\Delta\lambda_{\rm D}},
\label{vh}
\end{equation}
\begin{figure}
\centering
\includegraphics[width=10.0cm,height=11.0cm]{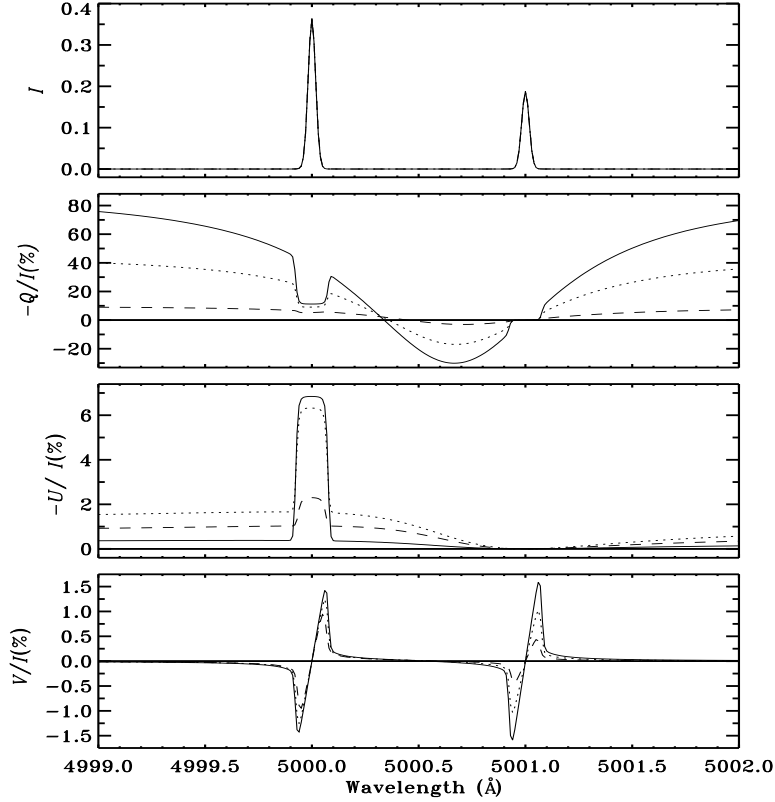}
\caption{Effect of collisions: The profiles of the intensity $I$ and the fractional
polarizations $-Q/I,-U/I$ and $V/I$ are plotted for a hypothetical
doublet at 5000\  \AA \ and 5001\  \AA \ with field
strength parameter $v_{\rm H}=0.004$. 
The coherence fractions used are $\gamma_{coh}=0.9$ (thick solid line), 
$\gamma_{coh}=0.5$ (dotted line), and $\gamma_{coh}=0.1$ (dashed line). 
The fine structure splitting is 1\ \AA. Single $90^{\circ}$ scattering 
is assumed at the extreme solar limb ($\mu=0$). 
The model parameters are $a_R=0.001$, $\vartheta_{B}=90^{\circ}$ and 
$\varphi_{B}=45^{\circ}$. The Doppler width $\Delta\lambda_{\rm D}=0.025$\ \AA.}
\label{gamebyr-ss}
\end{figure}
where $B$ is the magnetic field strength, $e_{0}$ is the charge of the
electron and $m$ its mass. $\Delta\lambda_{\rm D}$ is the Doppler
width and is assumed to be 0.025 \AA\ for both the lines. 
The radiative width of the upper state is parametrized as 
$a_{Rb}=\Gamma_{Rb}/(4\pi\Delta\nu_{\rm D})$. It is assumed to be the 
same for both the lines and is chosen to be 0.001. 
The radiative width $a_{Rb}$ is related to the total damping parameter through 
\begin{equation}
 a_b=a_{Rb}\Big[1 + \Big(\frac{\Gamma_{Ib}+\Gamma_{E}}{\Gamma_{Rb}}\Big)\Big].
\end{equation}
We assume the inelastic collision rate $\Gamma_{Ib}$ to be zero.

The depolarizing 
collisional rates $D^{(2)}=0.5\Gamma_{E}$, and $D^{(0)}=0$. { For 
simplicity we set $D^{(1)}=D^{(2)}$. However in general 
they can differ (for example, $D^{(1)}=0.43\Gamma_{E}$ and 
$D^{(2)}=0.38\Gamma_{E}$ according to \citet{berman69}). 
We have verified that {the} Stokes $V/I$ is insensitive 
to the values of $D^{(1)}$.}


The collisional effects are  built into the ${\bf R}$ matrix derived 
in Section~\ref{hz-tkq} through the branching ratios defined in Equations~(\ref{branch-b})
and (\ref{branch-a}). 
{The elastic collision rate is parametrized through the coherence fraction 
$\gamma_{coh}$  as 
$\gamma_{coh}=1/[1+(\Gamma_{E}/\overline{\Gamma}_{R}^{b{b^\prime}})].$
In the present paper we take $\Gamma_R$ to be the same for both
the upper fine structure states and $\gamma_{coh}$ to be the same
for all $J_bJ_{b^\prime}$ combinations.}
When $\gamma_{coh}=1$ the redistribution is entirely radiative 
(only ${\bf R}^{\rm II}$), whereas $\gamma_{coh}=0$ represents purely  
collisional redistribution (only ${\bf R}^{\rm III}$). We consider a range of 
values $\gamma_{coh} \in [1,0]$ to represent an arbitrary mix of 
${\bf R}^{\rm II}$ and ${\bf R}^{\rm III}$ type redistribution.

Figure~\ref{gamebyr-ss} shows the Stokes $(I, Q/I, U/I, V/I)$ spectra 
for a doublet. The polarization of the line at 
5001\ \AA\ is zero because its polarizability factor 
$W_2=0$. The collisions affect the 
wavelength domain outside the line core region 
of this line. But for the line at 5000\,\AA, the collisional effects are 
seen both in the line wings and the line core. In the line core the collisional 
effects compete with the Hanle effect and in the wings it is an interplay 
between the $J$-state interference effect and the collisional redistribution effect. 

The value $\gamma_{coh}=0.9$ corresponds to a mix with $90\%$ of ${\bf R}^{\rm II}$ 
and 10\% of ${\bf R}^{\rm III}$ (see the thick solid line in Figure~{\ref{gamebyr-ss}}). 
The profiles look similar to those for pure ${\bf R}^{\rm II}$ 
(see dotted line in Figure~3 of P1). However, in $Q/I$ there is a small 
depolarization, mainly in the wings, due to the presence of collisions. 
The core of the line at 5000\ \AA\ seems to be less affected than 
its wings. {The} $Q/I$ at the 5001\ \AA\ line remains zero.  
In the presence of elastic collisions, a small $U/I$ signal is generated in the wings
of the 5000\,\AA\ line. This non-zero 
$U/I$  wing polarization and the depolarization in the wings 
of $Q/I$ are induced by the elastic collisions in combination with the magnetic field
and can together be referred to as the `wing Hanle effect'. This effect arises because  
{the} elastic collisions can transfer the Hanle rotation (of the plane of
polarization) from the 
line core to the line wings before spontaneous de-excitation intervenes. In 
other words, in the presence of a small but significant elastic
collision rate the Hanle effect does not vanish in the line wings. If the elastic 
collision rate is large then {the collisions} completely depolarize the scattered
 radiation 
throughout the line profile. This effect has 
been discussed in detail in HZ2 for the case of a $J=0 \to 1 \to 0$
single scattering transition. However, these effects do not survive when the 
radiative transfer
effects with angle-averaged PRD are explicitly taken into account. 
The collisional redistribution process is more effective in
the case of angle-dependent PRD than in the case of angle-averaged PRD.
Using the domain-based PRD theory of \citet{vb97b} this effect 
was noticed even in the radiative transfer computations of \citet{knn02} (see also 
\citet{knn03}).
It remains as effective in a pair of interfering doublet lines as 
in the case of a single line.
However, in \citet{ms09} it was shown that {the wing Hanle} effect alone is 
insufficient to explain the observed wing signatures in the $Q/I$ and $U/I$ 
profiles of the Ca\,{\sc i} 4227\,\AA\ line.

As $\gamma_{coh}$ decreases to 0.5, which represents an equal mix of 
${\bf R}^{\rm II}$ and ${\bf R}^{\rm III}$, the values of $Q/I$ in the 
wings of both the lines are significantly reduced (see dotted line in 
Figure~{\ref{gamebyr-ss}}). The collisional effects are now seen even 
in the core of the 5000\,\AA\ line. This results in a decrease of the 
$Q/I$ and $U/I$ signals at the center of this line. The $J$-state 
interference signatures in $Q/I$ are also modified. 
When $\gamma_{coh}$ is further reduced to 0.1, the effects of 
${\bf R}^{\rm III}$ start to dominate over those of ${\bf R}^{\rm II}$ 
and also over the $J$-state interference effects (see dashed line 
in Figure~{\ref{gamebyr-ss}}). As a result the signatures of the 
$J$-state interference begin to fade away. {The} $Q/I$ and $U/I$ 
start to approach zero throughout the line profiles. As 
$\gamma_{coh}$ is further decreased to 0.0001 (not shown in the figure), 
the collisional effects (through ${\bf R}^{\rm III}$) completely dominate the 
scattering process. This situation corresponds to a regime of 
extremely large line broadening. As a result the amplitude 
of $I$ becomes much smaller compared to the other cases. Also, 
the $Q/I,U/I$, and $V/I$ approach zero level throughout the line profiles.

\subsection{Polarized line profiles formed due to 
multiple scattering in an atmospheric slab}
\label{iso-rt}
\begin{figure}
\centering
\includegraphics[width=10.0cm,height=8.0cm]{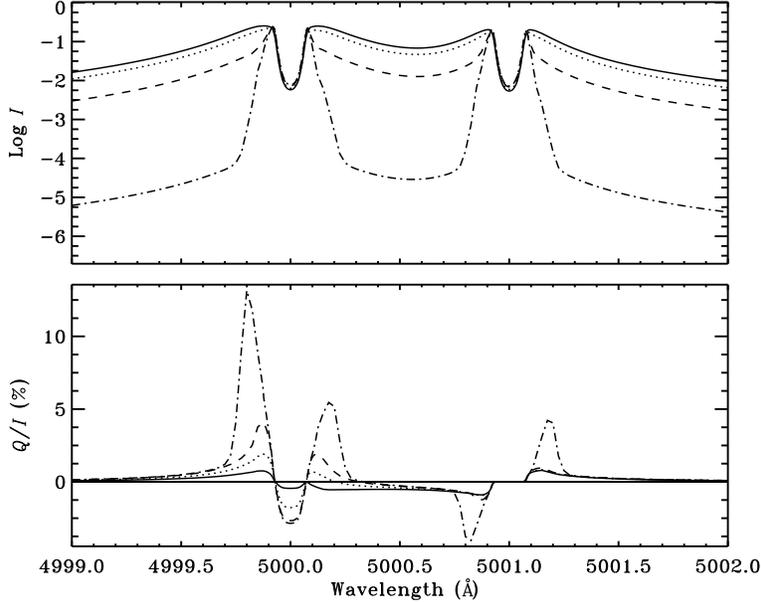}
\caption{Emergent Stokes profiles  at $\mu=0.047$ computed for a slab 
of optical thickness $T=2\times10^{4}$
in the absence of a background continuum. The other model parameters are
 $(a,\epsilon)=(10^{-3}, 10^{-4})$.
The magnetic field strength is set to zero. 
The coherence fraction is $\gamma_{coh}=0.1$ (solid line), 
$\gamma_{coh}=0.5$ (dotted line), $\gamma_{coh}=0.9$ (dashed line),
and $\gamma_{coh}=1$ (dot-dashed line).}
\label{gamebyr}
\end{figure}

In this section we present the emergent Stokes profiles computed
by solving the polarized radiative transfer equation
 for a two-term atom including the effects of $J$-state interference
and elastic collisions. For this we consider an isothermal constant
 property atmospheric slab
with a given optical thickness $T$. The slabs are
assumed to be self-emitting. 
The atmospheric model parameters used for the computations
are represented by $(T, a, \epsilon)$, where $a$ is the
damping parameter and $\epsilon$ is the thermalization 
parameter defined in Equation~(\ref{eps}) and the paragraph that follows.

The Planck function $B$ is taken as unity. 
The Doppler width for both {the} lines are assumed to be the same and equal to 0.025\ \AA. 
For more details on the structure of the atmospheric slabs and the model
parametrization we refer to P2.

\subsubsection{The non-magnetic case}

Figure~{\ref{gamebyr}} shows the emergent Stokes profiles which include the
effects of $J$-state interference, elastic collisions and radiative transfer
computed for a model atmosphere with parameters $T=2\times10^4, a=10^{-3}$ and
 $\epsilon=10^{-4}$
in the absence of a background continuum. In these profiles the  magnetic field 
is set to zero. 
Different line types represent different values of the coherence fraction
$\gamma_{coh}$. A range of values of $\gamma_{coh}\in[1,0]$ is considered.
As seen from the Figure~{\ref{gamebyr}}, a decrease in $\gamma_{coh}$ results in
a gradual decrease in $Q/I$ in the line core as well as in the 
PRD peaks of the 5000\ \AA\ line. The intensity profiles are also quite sensitive to the
effect of elastic collisions. 
As $\gamma_{coh}$ goes from 1 (pure $\bf R^{\rm II}$ case) to 0.1 
($\bf R^{\rm III}$ dominated
case), the self-reversed emission lines change over to nearly true absorption lines
(thick solid lines). 
In $Q/I$ the effects of collisions are confined only to the line core and
the near wing PRD peaks. Specifically, it is shown by \citet[][]{knn94} that 
the elastic collisions $D^{(2)}$ depolarizes the line core, and $\Gamma_{E}$ 
significantly depolarizes the line wing polarization (See \citet{mf92}).
The same conclusions are valid in the two-term atom model also. 
The interference region between the two lines seems to be 
less sensitive to the effect of elastic collisions. 

\begin{figure}
\centering
\includegraphics[width=10.0cm,height=8.0cm]{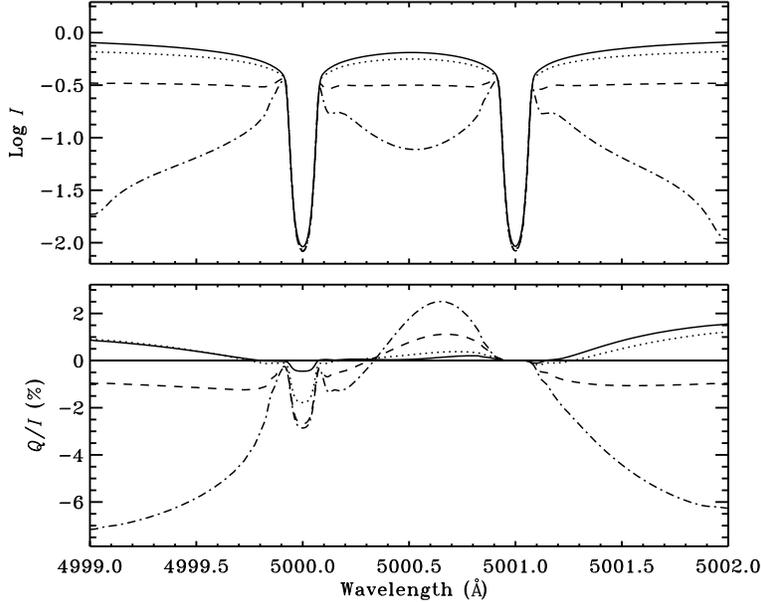}
\caption{Same as Figure~{\ref{gamebyr}} but computed for an optical 
thickness $T=2 \times 10^{8}$.}
\label{gamebyr-thick}
\end{figure}

However for larger optical depths, significant dependence on $\gamma_{coh}$
is exhibited in the wavelength region between the two lines. 
This can be seen in Figure~{\ref{gamebyr-thick}} which
shows the effect of elastic collisions in an optically
thick atmospheric slab in the absence of a magnetic field. The model parameters are 
the same as in Figure~{\ref{gamebyr}} but with $T=2\times 10^8$.
As $\gamma_{coh}$ decreases, the collisions take over the line
formation process. When $\gamma_{coh}=0.1$ (thick solid line), deep 
absorption lines are formed in $I$ with broad
wings. {The} $Q/I$ at the center of the 5000\,\AA \ line becomes very
 small like in Figure~{\ref{gamebyr}}.
The zero crossing point at 5000.3\ \AA\ remains the same for all the
values of $\gamma_{coh}$. 
In general a depolarization in $Q/I$ is seen throughout the line
profile because the radiative transfer effect is significant at all 
the frequencies. As expected, 
the $Q/I$ reaches zero very far in the wings of both the lines after 
exhibiting a wing maximum nearly 10\ \AA\ 
away from their line centers. 
The difference in behavior in the line core as well as in the line wings
 of the  $Q/I$ profiles
formed under $\bf R^{\rm II}$ dominated (dot-dashed line) and
 $\bf R^{\rm III}$ dominated (thick solid line) conditions
are better seen for the $T=2\times 10^{8}$ case when compared to the
$T=2 \times 10^{4}$ case. 

\subsubsection{The magnetic case}

\begin{figure}
\centering
\includegraphics[width=12.0cm,height=13.0cm]{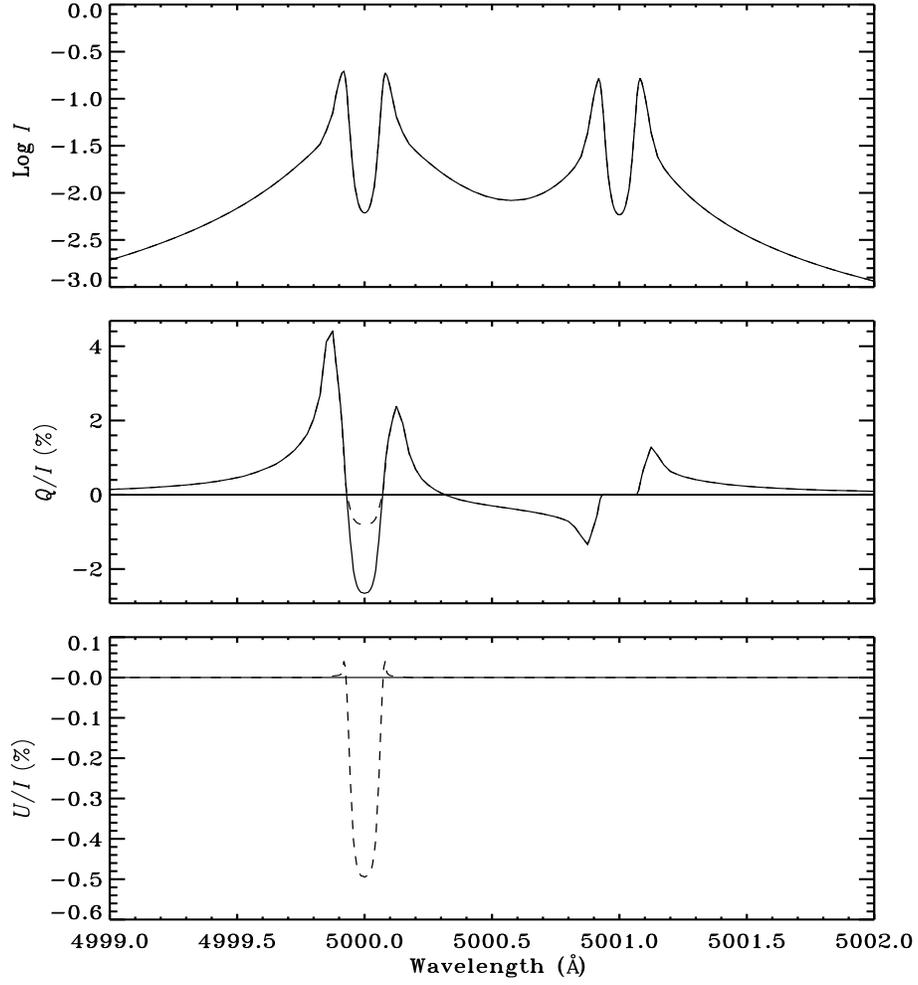}
\caption{Emergent Stokes profiles at $\mu=0.047$ computed for a magnetic 
field strength of
$v_{\rm H}=0$ (solid line) and $v_{\rm H}=0.004$ (dashed line) with a 
coherence fraction
 $\gamma_{coh}=0.9$ in the absence of a background continuum. 
The model parameters are $(T, a, \epsilon) = (2 \times 10^4, 10^{-3}, 10^{-4})$.}
\label{gamebyr-rt-mag}
\end{figure}

Figure~{\ref{gamebyr-rt-mag}} shows a comparison between the 
emergent Stokes profiles computed with (dashed line) and without (solid line)
 the presence of a weak magnetic field
including the effects of elastic collisions. 
The magnetic profiles are computed for a field strength of 
$v_{\rm H}=0.004$ with $\gamma_{coh}=0.9$. The model parameters are $T=2 \times 10^4$, 
$\epsilon=10^{-4}$, $a=10^{-3}$ in the absence of a background continuum. 
An external weak magnetic field (through the Hanle effect) affects the
 multiply scattered Stokes profiles
in a way similar to the singly scattered Stokes profiles. The Hanle effect
causes a depolarization in $Q/I$ at the center of the 5000\,\AA\ line
and also generates a $U/I$ signal at this line. We recall that these effects 
are not seen at
the 5001\,\AA \ line since its polarizability factor $W_2=0$. Like in the case
of single scattered profiles, the magnetic field effects are confined only
to the line core and the $J$-state interference signatures remain unaffected
by the magnetic field. Also as discussed earlier, the wing Hanle effect in
 $Q/I$ and $U/I$ which were seen 
in {the} case of single scattered profiles in Figure~{\ref{gamebyr-ss}}
now disappear due to the radiative transfer effects. 

\section{Conclusions}
\label{conclu}
{ In the present paper we have extended the theoretical framework for the
$J$-state interference for type-II redistribution developed in P1,
to include the effects of collisions (type-III 
redistribution). The collisional PRD matrix is derived in the
laboratory frame for a two-term atom with an unpolarized lower term and
in the presence of magnetic fields of arbitrary strengths. However, the treatment is 
restricted to the linear Zeeman regime for which the 
Zeeman splitting is much smaller than the fine-structure splitting. 
{The inelastic collisions coupling the upper term and the lower term
and also the inelastic collisions coupling the fine structure states
of the upper term are taken into account. However, the latter has been
treated approximately. The approximation involves
considering only the depolarizing
effects of the inelastic collisions but neglecting the 
polarization transfer rates between the fine structure states. 
The depolarization caused by the inelastic collisions
has the same type of consequences as the depolarization by elastic collisions.
Therefore we can merge both these effects into a common damping rate for the state $J_b$
and appropriately  redefine the branching ratios and the thermalization 
parameter for a two-term atom.
 A proper treatment of the inelastic collisions which
cause polarization transfer requires formulating and solving the 
polarized statistical equilibrium equations. This is outside 
the scope of the present paper. 
{The approximate treatment presented in this paper leads to 
slightly larger values of polarization in the line core as the 
inelastic collisions are not handled exactly. A treatment involving
the statistical equilibrium equation would yield correct values
of linear polarization. However in the line wings the formulation
presented here becomes accurate enough and would give the same 
result as a full treatment in terms of statistical equilibrium equation, including
PRD mechanism.}


The collisional frequency shift is inherently built into the 
redistribution matrix through the type-III redistribution function 
and the branching ratios. We discuss in detail the procedure of assigning the 
correct multipolar index $K$ to the collisional branching ratio
{and depolarizing elastic collision rate $D^{(K)}$}. This
procedure requires a detailed 
understanding of the role played by the multipolar index $K$ 
for both the atom and the radiation field.  We show how it becomes 
necessary to  introduce a quantum generalized profile function for the
case of a two-term atom in order to assign appropriate index $K$ to the
branching ratios {and to $D^{(K)}$. In general $D^{(K)}$ is defined for 
each of the fine structure components (by making it depend
on the quantum number $J_b$). 
{However in the present paper we assume it to be independent
of the $J$-quantum number.}

Examples of the Stokes profiles resulting from single $90$\textdegree \  
scattering are illustrated for different values of the coherence 
fraction $\gamma_{coh}$. The profiles look similar 
to the ones presented in P1, which were computed using collisionless redistribution 
(the case of pure ${\bf R}^{\rm II}$), except for the depolarization in 
the wings of the $Q/I$ profiles, and non-zero polarization in the wings 
of the $U/I$ profiles. This interesting feature, which we refer to as the 
wing Hanle effect, is discussed.

{The effects of collisions are discussed by incorporating the newly
derived collisional redistribution matrix in the polarized radiative transfer 
equation, in 
the simpler case of isothermal slab models.  The technique of
incorporating the Hanle redistribution matrix with the $J$-state interference
and collisions, into the polarized radiative transfer equation for a two-term atom
is presented. It is shown that the effects of elastic collisions in a two-term 
atom are similar
to those of the two-level atom case.} The redistribution matrices derived here
have been used in the interpretation of the quantum interference
signatures seen in the limb observations of the Cr\,{\sc i} triplet
in \citet{p3}. {For simplicity the inelastic collisions between the fine structure
states were neglected in that realistic modeling effort.}

With the present work we have further extended the theoretical tools
that are needed for modelling the various spectral structures 
arising due to the transitions between fine structure states
of an atom that have been observed in the Second Solar Spectrum 
so that they can be used to diagnose magnetic fields in regimes 
not accessible to the Zeeman effect. 

\section*{Acknowledgments}
M. Sampoorna is grateful to Drs. J. Trujillo Bueno and
 E. Landi Degl'Innocenti for useful discussions
on the density matrix approach.
The authors are grateful to Dr. V. Bommier for providing
programs to compute $R^{\rm III}$ function
of Hummer and a highly accurate code to evaluate
the corresponding angle-averaged functions.

\appendix
\section{Magnetic redistribution functions for type-III redistribution}
\label{appendix}
{In this appendix we present the expressions for the magnetic
redistribution functions of the type HH, HF, FH and FF
appearing in Equations~(\ref{h3mubmubprimemuf}) and (\ref{f3mubmubprimemuf}).
They are defined as follows
\begin{eqnarray}
R^{\rm III,\,HH}_{b_ma_m,{b^\prime}_mf_m}
(x_{b^\prime f},\,x^{\prime}_{ba},\,\Theta) &=& 
{\frac{1}{\pi^2 \sin \Theta}}
\int_{-\infty}^{+\infty}    du\,e^{-u^2} \left[{\frac{a_b}
{a_b^2+(v^\prime_{b_ma_m}-u)^2}}\right] \nonumber  \\ && \times 
H\left({\frac{a_{b^\prime}}{\sin\Theta}},
\, {\frac{v_{{b^\prime}_mf_m}}{\sin\Theta}}-
u\cot\Theta\right),
\label{rhh3}
\end{eqnarray}
\begin{eqnarray}
R^{\rm III,\,HF}_{b_ma_m,{b^\prime}_mf_m}
(x_{b^\prime f},\,x^{\prime}_{ba},\,\Theta) &=&
{\frac{1}{\pi^2 \sin \Theta}}
\int_{-\infty}^{+\infty}\,du\,e^{-u^2}\,   \left[{\frac{a_b}
{a_b^2+(v^{\prime}_{b_ma_m}-u)^2}}\right]\, \nonumber \\ && \times 
2F\left({\frac{a_{b^\prime}}{\sin\Theta}},
\, {\frac{v_{{b^\prime}_mf_m}}{\sin\Theta}}-
u\cot\Theta\right),
\label{rhf3}
\end{eqnarray}
\begin{eqnarray}
R^{\rm III,\,FH}_{b_ma_m,{b^\prime}_mf_m}
(x_{b^\prime f},\,x^{\prime}_{ba},\,\Theta) &=&
{\frac{1}{\pi^2 \sin \Theta}}
\int_{-\infty}^{+\infty}\,du\,e^{-u^2}\,                    
\left[{\frac{(v^{\prime}_{b_ma_m}-u)}
{a_b^2+(v^{\prime}_{b_ma_m}-u)^2}}\right]\,
\nonumber \\ && \times  H\left({\frac{a_{b^\prime}}{\sin\Theta}},
\, {\frac{v_{{b^\prime}_mf_m}} {\sin\Theta}}-
u\cot\Theta\right),
\label{rfh3}
\end{eqnarray}
and
\begin{eqnarray}
R^{\rm III,\,FF}_{b_ma_m,{b^\prime}_mf_m}
(x_{b^\prime f},\,x^{\prime}_{ba},\,\Theta)&=& 
{\frac{1}{\pi^2 \sin \Theta}}
\int_{-\infty}^{+\infty}\,du\,e^{-u^2}\,                    
\left[{\frac{(v^{\prime}_{b_ma_m}-u)}
{a_b^2+(v^{\prime}_{b_ma_m}-u)^2}}\right]\,
\nonumber \\ && \times 2F\left({\frac{a_{b^\prime}}{\sin\Theta}},
\, {\frac{v_{{b^\prime}_mf_m}} {\sin\Theta}}-u\cot\Theta\right).
\label{rff3}
\end{eqnarray}
In the above equations $H(a,x)$ and $F(a,x)$ are the 
Voigt and Faraday-Voigt functions (see Equation~(18) of P1 for their definition).
 $\Theta$ is the scattering angle (the angle between the incident and scattered rays; see 
Figure~1 of P1). The dimensionless quantities appearing in 
Equations~(\ref{rhh3}) to (\ref{rff3}) are given by 
\begin{eqnarray}
x_{ba}={\nu_{0ba}-\nu \over \Delta\nu_{\rm D}}; \quad 
a_b = {\frac{\gamma_{b}+\gamma_c}{4\pi\Delta \nu_{{\rm D}}}};\quad v_{b_ma_m}=
 x_{ba}+(g_b\mu_b-g_a\mu_a)\frac{\nu_L}{\Delta \nu_{{\rm D}}},
\label{nondim-quant1}
\end{eqnarray}
where $\nu_{0ba}$ is the line center frequency corresponding to
a $J_a\to J_b$ 
transition in the absence of magnetic fields, 
$a_b$ is the damping parameter
of the excited state $b$, 
and ${\Delta \nu_{{\rm D}}}$ is the Doppler width.
In the limit of a two-level atom (obtained by setting 
$J_b=J_b'$  and $J_a=J_f$) and in the absence of a magnetic 
field the $R^{\rm III,\,HH}$ and $R^{\rm II,\,H}$ (defined in P1)
reduce to the Hummer's $R^{\rm III}$ and $R^{\rm II}$ functions respectively
(see also HZ1). 

The angle-averaged analogues of Equations~(\ref{rhh3})-(\ref{rff3})
are obtained through
\begin{equation}
 R^{\rm III,\,XY}_{b_ma_m,{b^\prime}_mf_m}
(x_{b^\prime f},\,x^{\prime}_{ba})=
\frac{1}{2}\int_0^\pi  R^{\rm III,\,XY}_{b_ma_m,{b^\prime}_mf_m}
(x_{b^\prime f},\,x^{\prime}_{ba},\,\Theta)\ \sin\Theta\  d\Theta,
\label{aa-rf}
\end{equation}
(see Equations~(103) and (104) of \citet{vb97b} and 
Equations~(30) and (31) of \citet{ms08}), where X and Y 
stand for H and/or F. A similar expression can be used
for computing angle-averaged analogues of the type-II functions.

\bibliographystyle{model3-num-names}

\end{document}